\documentclass[10pt,journal,comsoc]{IEEEtran}

\usepackage{amsmath,amsfonts,amssymb,amsthm}
\usepackage{algorithmic}
\usepackage[algo2e, ruled, vlined, linesnumbered]{algorithm2e}

\usepackage{array}
\usepackage{textcomp}
\usepackage{stfloats}
\usepackage{verbatim}
\usepackage{graphicx}
\usepackage{booktabs}

\usepackage{caption}
\usepackage{subcaption}
\usepackage{multirow}
\usepackage{xcolor}
\usepackage{eso-pic}

\makeatletter
\renewcommand\p@subfigure{\thefigure} 
\makeatother

\def\BibTeX{{\rm B\kern-.05em{\sc i\kern-.025em b}\kern-.08em
    T\kern-.1667em\lower.7ex\hbox{E}\kern-.125emX}}

\SetKwProg{Fn}{Function}{}{end}

\SetKwInput{KwInit}{Init}
\SetAlgoSkip{}

\newcolumntype{M}[1]{>{\centering\arraybackslash}m{#1}}

\DeclareSymbolFont{symbolsC}{U}{txsyc}{m}{n}
\DeclareMathSymbol{\notniFromTxfonts}{\mathrel}{symbolsC}{61}

\usepackage{titlesec}

\titlespacing{\section}{0pt}{6pt plus .2ex minus .2ex}{0.5ex plus .2ex}
\titlespacing{\subsection}{0pt}{6pt plus .2ex minus .2ex}{0.5ex plus .2ex}

\newtheoremstyle{remarkstyle}
  {3pt}                % 上方間距
  {3pt}                % 下方間距
  {}                   % 內文體（通常用斜體，留空則為正體）
  {}                   % 縮排量
  {\bfseries}          % 標題字體（加粗）
  {.}                  % 標題後的標點
  {.5em}               % 標題後的間距
  {}                   % 標題自定義（通常留空）

\theoremstyle{remarkstyle}
\newtheorem{definition}{Definition}
\newtheorem{theorem}{Theorem}
\newtheorem{lemma}{Lemma}
\newtheorem{remark}{Remark} 
\newtheorem{proposition}{Proposition} 
\newtheorem{corollary}{Corollary}

\title{Semantic-aware Digital Twins for Resilient Service Coverage in Low-Altitude Economy Networks with Intermittent Connectivity}

\author{Chuan-Chi~Lai,~\IEEEmembership{Member,~IEEE}, Ang-Hsun~Tsai,~\IEEEmembership{Member,~IEEE}, and Zhu Han,~\IEEEmembership{Fellow,~IEEE}
    \IEEEcompsocitemizethanks{
        \IEEEcompsocthanksitem{
            This work was supported by the National Science and Technology Council, Taiwan, under Grant Nos. NSTC 114-2221-E-194-062-, NSTC 115-2221-E-194-042-MY2, and NSTC 115-2221-E-035-050-. This work was also partially supported by the Advanced Institute of Manufacturing with High-tech Innovations (AIM-HI) from the Featured Areas Research Center Program within the framework of the Higher Education Sprout Project by the Ministry of Education (MOE) in Taiwan. \emph{(Corresponding author: Chuan-Chi~Lai.)}}
        \IEEEcompsocthanksitem{
            Chuan-Chi~Lai is with the Department of Communications Engineering, National Chung Cheng University, Minxiong Township, Chiayi County 621301, Taiwan, and also with the Advanced Institute of Manufacturing with High-tech Innovations (AIM-HI), National Chung Cheng University, Minxiong Township, Chiayi County 621301, Taiwan (e-mail: chuanclai@ccu.edu.tw).}
        \IEEEcompsocthanksitem{
            Ang-Hsun~Tsai is with the Department of Communications Engineering, Feng Chia University, Taichung 407102, Taiwan (e-mail: ahtsai@fcu.edu.tw).}
        \IEEEcompsocthanksitem{
            Zhu Han is with the Department of Electrical and Computer Engineering, University of Houston, Houston, TX 77204, USA (e-mail: zhan2@uh.edu).}       
    }
}

\IEEEtitleabstractindextext{%
\begin{abstract}
    Application-aware Digital Twin (DT) networks require reliable closed-loop intelligence, yet intermittent aerial connectivity makes continuous synchronization between UAV swarms and their digital representations impractical. Prolonged communication voids therefore cause increasing world-model mismatch and degrade autonomous coordination. To address this challenge, we propose the Embodied Proactive Inference and Coordination (EPIC) framework based on Federated Multi-Agent Reinforcement Learning (FMARL). EPIC couples open-loop state inference with closed-loop federated realignment: Feature-based Spatio-Temporal Inference (F-STI) reconstructs latent state evolution during communication voids, while asymmetric federated calibration realigns distributed semantic world models when connectivity becomes available. A Value of Information (VoI) mechanism further prioritizes task-critical synchronization opportunities to reduce redundant signaling. The resulting decentralized coordination is characterized by a state-conditional congestion game, providing an equilibrium reference for deriving an $\epsilon$-Nash robustness bound when state uncertainty remains bounded under the stated regularity conditions. Simulations demonstrate a nominal 66.7\% reduction in scheduled uplink frequency while sustaining service coverage and rapidly recovering state alignment after prolonged communication voids.
\end{abstract}

}

\begin{document}

% --- 縮減數學公式前後的空白 ---
\setlength{\abovedisplayskip}{3pt}
\setlength{\belowdisplayskip}{3pt}
\setlength{\abovedisplayshortskip}{3pt}
\setlength{\belowdisplayshortskip}{3pt}
% --- 縮減圖片/表格與內文之間的空白 ---
\setlength{\textfloatsep}{6pt plus 2pt minus 2pt}
\setlength{\floatsep}{6pt plus 2pt minus 2pt}
\setlength{\intextsep}{6pt plus 2pt minus 2pt}

\AddToShipoutPictureBG*{% 帶星號表示僅套用至當前（第一）頁
    \AtPageUpperLeft{%
        \raisebox{-1.3cm}[\height][\depth]{% 控制距離頁面頂端的垂直距離
            \makebox[\paperwidth][c]{% 以全頁寬度置中
                \parbox{.9\textwidth}{% 設定段落寬度為本文寬度，並支援自動折行
                    \centering
                    \footnotesize \copyright~2026 IEEE. Personal use of this material is permitted. Permission from IEEE must be obtained for all other uses, in any current or future media, including reprinting/republishing this material for advertising or promotional purposes, creating new collective works, for resale or redistribution to servers or lists, or reuse of any copyrighted component of this work in other works.
                }%
            }%
        }%
    }%
}

\maketitle
\IEEEdisplaynontitleabstractindextext
\IEEEpeerreviewmaketitle

\section{Introduction}
\label{sec:intro}
\IEEEPARstart{T}{he} rapid evolution of the \textit{Low-Altitude Economy} (LAE) drives large-scale \textit{Unmanned Aerial Vehicle} (UAV) deployments for mission-critical tasks such as environmental monitoring and emergency logistics \cite{11131292, Shamsoshoara2021}. Ensuring resilient coordination within these networks is fundamental to reliable service delivery \cite{11194218,11416901}. However, conventional multi-agent frameworks rely heavily on frequent information exchange, which becomes increasingly difficult in highly dynamic aerial networks with limited spectral resources and volatile wireless links \cite{11269020}.

In communication-constrained aerial environments, intermittent connectivity is often unavoidable due to dynamic topology, stochastic link conditions, and the need to limit signaling overhead. While the \textit{Age of Information} (AoI) is widely used to characterize temporal freshness \cite{11194218, 10310169}, it does not directly capture the discrepancy between ground-truth physical states and their remote digital representations. Consequently, the \textit{Age of Incorrect Information} (AoII) provides a more informative measure of mission-critical state degradation \cite{Tong2023_AoII_DT}. Under prolonged signaling silence, however, conventional reactive \textit{Multi-Agent Reinforcement Learning} (MARL) policies must operate with increasingly stale information, which can lead to coordination drift, redundant coverage, or misdirected service allocation \cite{Yu2024Effective, Chen2020Track}.

These observations expose a fundamental operational dilemma in resilient swarm coordination. During communication voids, autonomous UAVs cannot rely on fresh peer information and must instead infer the evolution of the unobserved world state from locally maintained information. Such open-loop inference, however, inevitably accumulates model mismatch and kinematic prediction error over time. Conversely, continuously enforcing tight closed-loop synchronization is incompatible with the spectral and connectivity constraints that create these communication voids. Hence, resilient coordination requires a complementary two-phase architecture: \emph{proactive open-loop inference during communication voids, coupled with opportunistic closed-loop realignment when connectivity becomes available}.

Motivated by this design principle, we propose the \textit{Embodied Proactive Inference and Coordination} (EPIC) framework. EPIC systematically bridges the open-loop and closed-loop phases within a unified \textit{Federated Multi-Agent Reinforcement Learning} (FMARL) architecture. During communication voids, each agent employs a \textit{Feature-based Spatio-Temporal Inference} (F-STI) operator with damped kinematic priors to extrapolate peer trajectory evolution and maintain its local semantic world model. When synchronization opportunities arise, an asymmetric federated calibration mechanism realigns distributed world-model representations toward a shared model reference. A mission-aware \textit{Value of Information} (VoI) mechanism further determines whether the accumulated uncertainty and task relevance justify synchronization, thereby coupling inference, communication, and coordination in an application-aware closed-loop architecture.

The main contributions of this paper to the field of networked intelligence are summarized as follows:
\begin{itemize}
    \item \textit{Open-loop/closed-loop coordination paradigm for intermittent connectivity}: We establish a unified aerial coordination architecture that systematically couples open-loop state inference during communication voids with closed-loop federated realignment upon connectivity recovery. This formulation explicitly addresses the fundamental trade-off between decentralized operational autonomy under stale information and synchronization fidelity under stringent signaling limits.
    
    \item \textit{Proactive semantic world-model synchronization}: We develop the EPIC framework by synergizing the F-STI operator with an asymmetric federated calibration protocol. F-STI provides continuous open-loop trajectory reconstruction during signaling gaps, while federated aggregation selectively realigns distributed world-model representations when synchronization becomes available. Guided by an opportunistic VoI trigger, EPIC achieves a nominal 66.7\% reduction in scheduled uplink frequency while maintaining high coordination consistency.

    \item \textit{Theoretical characterization and empirical resilience under extreme communication voids}: We decouple the system-level cooperative coverage objective from localized congestion-based incentives and characterize the resulting interaction as a state-conditional congestion game. Its potential structure provides an equilibrium reference, from which we derive an $\epsilon$-Nash robustness bound when state uncertainty remains bounded under the stated regularity conditions. Extensive simulations further indicate that EPIC sustains resilient service coverage over extended 30-step communication voids, demonstrating favorable operational robustness across the evaluated swarm densities and task environments.
\end{itemize}

The remainder of this paper is organized as follows. Section~\ref{sec:related_work} reviews the related literature, Section~\ref{sec:system_model} presents the system model and problem formulation, Section~\ref{sec:EPIC_framework} describes the EPIC framework, Section~\ref{sec:theoretical_analysis} conducts the theoretical analysis, Section~\ref{sec:simulation_results} presents the simulation results, and finally Section~\ref{sec:conclusion} concludes the paper.

\section{Related Work}
\label{sec:related_work}

UAV coordination in dynamic aerial networks has been studied from multiple perspectives, including robust control, edge-assisted coordination, energy-efficient trajectory planning, and adaptive learning. Robust control has been used to mitigate adversarial disturbances \cite{10584430}, while edge-assisted access supports scalable mission execution \cite{Pham2022}. Recent studies have also explored graph-based multi-agent learning for UAV coordination \cite{Tang2025}. In the Low-Altitude Economy (LAE), energy-efficient trajectory optimization has been investigated for dynamic service demands \cite{11194218}, alongside multi-modal adaptive policies for heterogeneous tasks \cite{Wang2026_MultiModal}. Network resilience has further been studied from integrated sensing, communication, and computation perspectives \cite{11368666}, as well as through topology-aware mechanisms \cite{lai2026resilient} and proactive spatio-temporal state reconstruction \cite{LaiWCL2026}. These studies highlight the need for reliable coordination and state awareness under dynamic and intermittent network conditions.

With localized communication and sensing, multi-agent coordination can be formulated as a \textit{Decentralized Partially Observable Markov Decision Process} (Dec-POMDP) \cite{Bernstein2002, oliehoek2016concise}. Communication-efficient coordination has been studied through selective information exchange, including weight-based communication scheduling \cite{kim2019learning} and sparse dual-attention architectures \cite{Chen2026_SparseAttn}, and importance-aware semantic feedback that prioritizes task-relevant information for transmission \cite{zhou2026optimization}. Bayesian forecasting provides a principled approach to latent-state prediction \cite{west1997bayesian}, while state-tracking methods have also been integrated into multi-agent decision systems \cite{Chen2020Track}. Under prolonged communication gaps, agents increasingly rely on local information and prediction, motivating proactive state inference.

Information freshness is another important consideration in communication-constrained coordination. The AoI is widely used to characterize the temporal freshness of networked information \cite{Yates2021, Hu2025_AoI_LAE, Chen2026_STA_DRL}. For remote decision-making and estimation, however, freshness alone does not capture the discrepancy between physical states and their remote representations. The AoII therefore provides a complementary measure that accounts for both information staleness and state estimation accuracy \cite{maatouk2020age, Tong2023_AoII_DT}. Building on this perspective, EPIC uses a mission-aware VoI mechanism to quantify task uncertainty and determine when synchronization is sufficiently valuable during intermittent connectivity.

\textit{Digital Twin} (DT) technology provides a framework for maintaining digital representations of physical systems and supporting closed-loop monitoring, prediction, and control \cite{10938590, Zhou2026DT, Luo2024, 11204513}. In wireless and aerial networks, distributed DTs must reconcile local observations, prediction, and communication constraints, making synchronization latency and bandwidth important considerations \cite{Li2025_DelayAware}. Recent studies have explored task-oriented and anomaly-aware DT operation, including SEMADT-RL, which combines state estimation and reinforcement learning for adaptive DT management \cite{Hazarika2025_SEMADT_RL}. Proactive spatio-temporal extrapolation has also been investigated to sustain state awareness when direct observations are unavailable \cite{LaiWCL2026}. These studies motivate task-oriented representations that retain useful state information between synchronization opportunities. In EPIC, this role is fulfilled by the Embodied Semantic World Model (ESWM) for proactive state reconstruction, information-aware synchronization, and decentralized coordination.

\textit{Federated Learning} (FL) provides a distributed mechanism for coordinating model knowledge without centrally transmitting raw local data \cite{Li2020FedAvg, 11193784, Du2024}. Wireless and multi-agent applications have used FL to reduce direct data exchange, while adaptive federated mechanisms address resource and participation heterogeneity \cite{Wang2019Adaptive}. For intermittently connected UAV swarms, however, federated synchronization must also accommodate heterogeneous communication opportunities and evolving local models. EPIC therefore adopts an asymmetric federated calibration mechanism that opportunistically aggregates model information when synchronization is available. Combined with F-STI-based proactive inference, it forms a complementary operating cycle: local agents maintain task-relevant state estimates during communication silences, while federated synchronization realigns distributed representations when connectivity becomes available. This integration closes the gap between proactive state maintenance, communication-aware synchronization, and decentralized coordination under intermittent connectivity within an application-aware closed-loop Digital Twin framework.

\section{System Model and Problem Formulation}
\label{sec:system_model}
\begin{figure}[!t]
\centering
    \includegraphics[width=.78\linewidth]{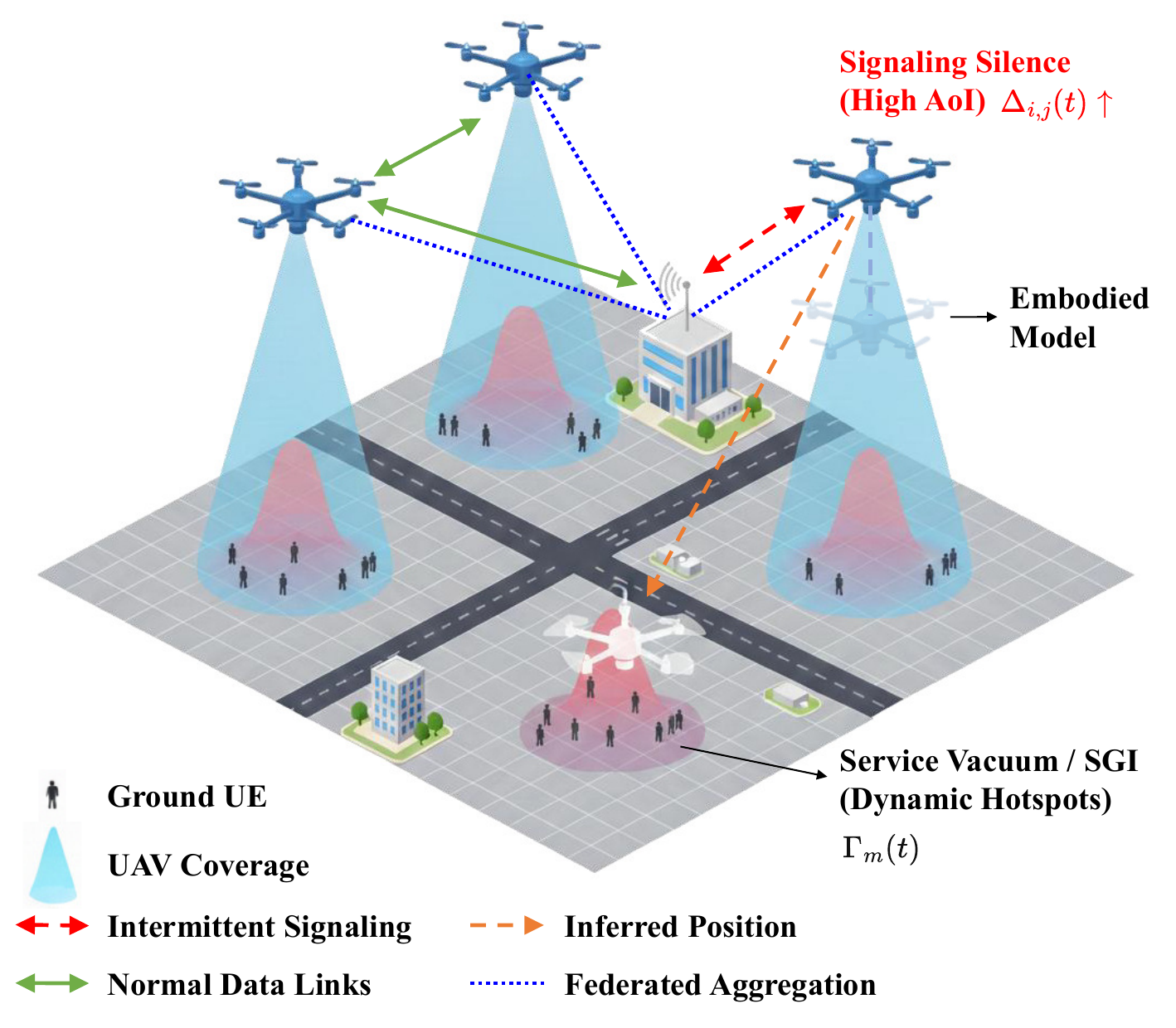}
    \caption{System architecture of the resilient LAE network. The swarm operates over dynamic mission hotspots. Under intermittent connectivity, agents experience prolonged silences leading to elevated information age, denoted by $\Delta_{i,j}(t)$. The EPIC framework enables agents to use onboard F-STI to reconstruct peer states during signaling silences and fill service gaps.}
    \label{fig:system_scenario}
\end{figure}

We mathematically characterize the UAV swarm environment within the LAE paradigm, explicitly capturing the multi-layer coupling between mission-level priority and network-layer information freshness. This coupling is quantified as the VoI, serving as a mission-aware proxy for the AoII to trigger proactive coordination, as illustrated in Fig.~\ref{fig:system_scenario}.

\subsection{Dynamic Mission Environment and Service Gap}
Consider a mission airspace $\mathcal{L} \times \mathcal{L}$ where a swarm of $N$ autonomous UAVs, $\mathcal{U} = \{u_1, \dots, u_N\}$, provides persistent service over $M$ discretized grids $\mathcal{G} = \{g_m\}_{m=1}^M$. To capture the spatiotemporal volatility of service demands across these grids \cite{Chen2026_STA_DRL}, we construct an application-specific \textit{Dynamic Gaussian-Mixture demand Model} (DGMM) inspired by~\cite{stauffer1999adaptive}:
\begin{equation}\label{eq:weight_map}
    w_m(t) = \max_{k \in \{1 \dots K\}} \left( A_k \exp \left( -\frac{\|\mathbf{g}_m - \boldsymbol{\mu}_k(t)\|^2}{2\sigma_k^2} \right) \right) + \epsilon,
\end{equation}
where $A_k$ denotes the maximum priority of the $k$-th dynamic hotspot out of $K$ total hotspots. Here, $\mathbf{g}_m \in \mathbb{R}^2$ is the fixed grid coordinate, $\boldsymbol{\mu}_k(t)$ is the time-varying centroid, $\sigma_k$ defines the spatial spread, and $\epsilon$ is a baseline constant.

Driven by these dynamic weights, we evaluate the physical coverage provided by the swarm. Let $\chi_{i,m}(t) \in \{0,1\}$ denote the binary coverage status, where $\chi_{i,m}(t) = 1$ if the Euclidean distance between UAV $u_i$'s position $\mathbf{p}_{i,t}$ and $\mathbf{g}_m$ is within the effective radius $R_{\mathrm{cov}}$. To quantify the accumulated service deficit at grid $g_m$, we define the \textit{Service Gap Intensity} (SGI), denoted by $\Gamma_m(t)$, which evolves recursively as
\begin{equation}\label{eq:SGI_evolution}
    \Gamma_m(t)
    =
    \mathrm{clip}
    \left(
    \Gamma_m(t-1) + \Delta \Gamma_m(t),
    \, 0,
    \, \Gamma_{\max}
    \right),
\end{equation}
where the instantaneous SGI variation is given by
\begin{equation}\label{eq:delta_SGI}
    \Delta\Gamma_m(t)
    =
    \delta_{\mathrm{age}} w_m(t)
    -
    \delta_{\mathrm{rec}}
    \cdot
    \mathbb{1}
    \left\{
    \exists i \in \mathcal{U}_{\mathrm{act}} :
    \chi_{i,m}(t) = 1
    \right\}.
\end{equation}
Here, $\mathcal{U}_{\mathrm{act}}\subseteq\mathcal{U}$ denotes the set of currently active, non-failed UAVs, while $\delta_{\mathrm{age}}$ and $\delta_{\mathrm{rec}}$ denote the service-deficit aging and recovery rates, respectively.

While the SGI captures localized service deficits, the overall mission fulfillment is evaluated by the \textit{Weighted Coverage Efficiency} (WCE) score:
\begin{equation}\label{eq:wce_final}
    \eta_{\mathrm{wce}}(t) = \sum_{m \in \mathcal{G}} \left( w_m(t) \cdot \mathbb{1} \left\{ \exists i \in \mathcal{U}_{\mathrm{act}}: \chi_{i,m}(t)=1 \right\} \right)^{\alpha_w},
\end{equation}
where $\alpha_w \ge 1$ is the non-linear WCE evaluation exponent. Conversely, the SGI provides a task-oriented proxy for unfulfilled service demand, whose accumulation during signaling silences informs the subsequent VoI assessment.

\subsection{Communication and Propagation Model}
We adopt a probabilistic \textit{Air-to-Ground} (A2G) propagation model \cite{AlHourani2014}. The Reference Signal Received Power (RSRP), $P_{i,k}(t)$, received by agent $u_i$ from beacon $\boldsymbol{\mu}_k(t)$ is:
\begin{equation}\label{eq:rsrp_model}
    P_{i,k}(t) = P_{\mathrm{tx}} - L_{i,k}(t) - \xi,
\end{equation}
where $P_{\mathrm{tx}}$ is the transmit power and $\xi \sim \mathcal{N}(0, \sigma^2_{\mathrm{shadow}})$ represents the shadowing margin. The path loss $L_{i,k}(t)$ depends on the Line-of-Sight (LoS) probability $P_{\mathrm{LoS}}$:
\begin{equation}\label{eq:Plos}
    P_{\mathrm{LoS}}(\vartheta_{i,k}) = \frac{1}{1 + a \exp(-b [\vartheta_{i,k} - a])},
\end{equation}
where $a,b$ are environmental constants, $\vartheta_{i,k}(t)$ is the elevation angle, $H$ is the fixed UAV altitude, and $d_{i,k}^{\mathrm{hor}}(t) =\|\mathbf p_{i,t}-\boldsymbol\mu_k(t)\|$ denotes the horizontal distance. The corresponding 3-D slant distance used for channel-gain evaluation is $d_{i,k}^{\mathrm{3D}}(t)=\sqrt{\left(d_{i,k}^{\mathrm{hor}}(t)\right)^2+H^2}$, and $\vartheta_{i,k}(t)=\frac{180}{\pi}\arcsin\left(\frac{H}{d_{i,k}^{\mathrm{3D}}(t)}
\right)$. The average path loss $L_{i,k}(t)$ is given by:
\begin{equation}\label{eq:path_loss}
    L_{i,k}(t) = 20\log_{10}\left(\frac{4\pi f_c d_{i,k}^{\mathrm{3D}}(t)}{c}\right) + P_{\mathrm{LoS}} \eta_{\mathrm{LoS}} + (1 - P_{\mathrm{LoS}}) \eta_{\mathrm{NLoS}},
\end{equation}
where $f_c$ is the carrier frequency, $c$ is the speed of light, and $\eta_{\mathrm{LoS}}$, $\eta_{\mathrm{NLoS}}$ are the attenuation factors for LoS and Non-LoS links.

To characterize the communication-state evolution in \eqref{eq:comm_cons_sinr}, we additionally define the \textit{Air-to-Air} (A2A) propagation model for the inter-agent signaling links.
Given that the UAVs operate at similar altitudes, the A2A channel is predominantly characterized by LoS conditions. The received power $P_{i,j}^{\mathrm{rx}}(t)$ at agent $u_i$ from agent $u_j$ is given by:
\begin{equation}\label{eq:rsrp_a2a}
    P_{i,j}^{\mathrm{rx}}(t) = P_{\mathrm{tx}} - L_{i,j}^{\mathrm{A2A}}(t) - \xi_{\mathrm{A2A}},
\end{equation}
where $\xi_{\mathrm{A2A}}$ represents the small-scale fading margin, and $L_{i,j}^{\mathrm{A2A}}(t)$ is the free-space path loss defined as:
\begin{equation}\label{eq:path_loss_a2a}
    L_{i,j}^{\mathrm{A2A}}(t) = 20 \log_{10} \left( \frac{4\pi f_c d_{i,j}(t)}{c} \right),
\end{equation}
where $d_{i,j}(t) = \|\mathbf{p}_{i,t} - \mathbf{p}_{j,t}\|$ is the Euclidean distance between the two agents. Consequently, the \textit{Signal-to-Interference-plus-Noise Ratio} (SINR) for the coordination link is evaluated as:
\begin{equation}\label{eq:sinr_a2a}
    \mathrm{SINR}_{i,j}(t) = \frac{10^{P_{i,j}^{\mathrm{rx}}(t)/10}}{\sum_{u \in \mathcal{U} \setminus \{i,j\}} 10^{P_{i,u}^{\mathrm{rx}}(t)/10} + 10^{N_0/10}},
\end{equation}
where $N_0$ denotes the background noise power in $\mathrm{dBm}$.

Similarly, for the UAV-to-infrastructure (A2I) uplink signaling link between agent $u_i$ and the edge base station (BS), the received power is modeled as:
\begin{equation}\label{eq:rsrp_a2i}
    P_{i,\mathrm{BS}}^{\mathrm{rx}}(t) = P_{\mathrm{tx}} - L_{i,\mathrm{BS}}(t) - \xi_{i,\mathrm{BS}},
\end{equation}
where the path loss $L_{i,\mathrm{BS}}(t)$ follows the elevation-dependent A2G model in \eqref{eq:Plos}--\eqref{eq:path_loss}. Under co-channel uplink transmissions from active peer UAVs, the corresponding uplink SINR is evaluated as:
\begin{equation}\label{eq:sinr_a2i}
    \mathrm{SINR}_{i,\mathrm{BS}}(t) = \frac{10^{P_{i,\mathrm{BS}}^{\mathrm{rx}}(t)/10}}{\sum_{u \in \mathcal{U} \setminus \{i\}} 10^{P_{u,\mathrm{BS}}^{\mathrm{rx}}(t)/10} + 10^{N_0/10}}.
\end{equation}

\noindent
\textit{Scope of the Communication Model:} The communication model is intended to characterize the availability and intermittency of coordination links rather than to jointly optimize physical-layer radio resources. Specifically, transmit powers and carrier frequency are treated as fixed system parameters, while propagation loss and co-channel interference determine the instantaneous SINR and hence the availability of A2A coordination and UAV-to-infrastructure signaling. Accordingly, EPIC focuses on semantic synchronization and coordination under these time-varying connectivity conditions, rather than jointly allocating spectrum, bandwidth, or transmit power.

\subsection{Problem Formulation}
\label{subsec:problem_formulation}

The global mission objective is to minimize the cumulative SGI and kinetic control effort subject to the UAV kinematic and safety constraints. The idealized optimization problem $\mathcal{P}_0$ is formulated as
\begin{align}
    \min_{\{\mathbf{a}_{i,t}\}} \quad
    & \mathcal{J}
    =
    \mathbb{E}
    \left[
    \sum_{t=0}^{T}
    \gamma^t
    \left(
    \sum_{m \in \mathcal{G}} \Gamma_m(t)
    +
    \sum_{i \in \mathcal{U}}
    \eta_{\mathrm{ops}}
    \|\mathbf{a}_{i,t}\|^2
    \right)
    \right]
    \label{eq:obj_main}\\
    \text{s.t.} \quad
    & \mathbf{p}_{i,t+1}
    =
    \mathbf{p}_{i,t}
    +
    \mathbf{v}_{i,t}\delta t,
    \quad \forall i,t,
    \label{eq:pos_update}\\
    &
    \mathbf{v}_{i,t}
    =
    \alpha \mathbf{a}_{i,t} v_{\mathrm{gain}}
    +
    (1-\alpha)\mathbf{v}_{i,t-1},
    \quad \forall i,t,
    \label{eq:inertia_dynamics}\\
    &
    \|\mathbf{v}_{i,t}\|
    \le v_{\max},
    \quad
    \|\mathbf{p}_{i,t}-\mathbf{p}_{j,t}\|
    \ge d_{\mathrm{safe}},
    \quad \forall i\neq j,t.
    \label{eq:physical_cons}
\end{align}

Here, $\mathcal{J}$ is the expected cumulative cost, $\gamma \in [0,1)$ is the discount factor, and $\eta_{\mathrm{ops}}$ weights the kinetic control effort. The vector $\mathbf{a}_{i,t}\in\mathbb{R}^2$ denotes the normalized control action. The kinematics in \eqref{eq:inertia_dynamics} are governed by the damping factor $\alpha\in[0,1]$, the velocity-retention factor $(1-\alpha)$, and the action-scaling factor $v_{\mathrm{gain}}$. The physical constraints in \eqref{eq:physical_cons} impose the maximum speed $v_{\max}$ and the pairwise safety-separation threshold $d_{\mathrm{safe}}$. In decentralized reinforcement learning execution under partial observability, the latter is incorporated through a soft reward penalty weighted by $\lambda_{\mathrm{col}}$.

Inter-agent information staleness is modeled separately as a communication-induced state variable rather than an optimization constraint. Specifically, the information age $\Delta_{i,j}(t)$ evolves according to
\begin{equation}
\label{eq:comm_cons_sinr}
    \Delta_{i,j}(t)=\begin{cases}
        0,
        & \text{if } \mathrm{SINR}_{i,j}(t) \ge \gamma_{\mathrm{th}},\\
        \min\!\left(\Delta_{i,j}(t-1)+1,\,T_{\mathrm{up}}\right),
        & \text{otherwise}.
    \end{cases}
\end{equation}
Accordingly, a successful coordination update resets $\Delta_{i,j}(t)$ to zero, whereas an unavailable link causes the state information to become progressively stale, up to the maximum outage horizon $T_{\mathrm{up}}$.

To accurately interpret the system dynamics under intermittent connectivity, it is critical to distinguish between the inter-agent \textit{coordination plane} and the \textit{service plane} provided to the ground hotspots. The communication-state update in \eqref{eq:comm_cons_sinr} governs information staleness on the coordination plane, where communication silences cause the information age to accumulate. In traditional reactive multi-agent systems, such coordination intermittency directly translates to severe service degradation, as agents fail to maintain spatial formations. However, EPIC is explicitly designed to reduce the impact of coordination intermittency on the service plane. By utilizing onboard state-inference mechanisms to sustain proactive local extrapolations, the agents can maintain high service quality, quantified by the WCE score in \eqref{eq:wce_final}, even when coordination links are severely disrupted. This design reduces the impact of communication volatility on ground service coverage, justifying the resilience of the service delivery.

\color{black}

\subsection{Problem Intractability and Value of Information}
\label{subsec:complexity_analysis}
The optimization problem $\mathcal{P}_0$ is non-convex and contains a combinatorial multi-agent coverage subproblem. Under the intermittent and partial observations induced by \eqref{eq:comm_cons_sinr}, the sequential coordination problem can be represented as a Decentralized Partially Observable Markov Decision Process (Dec-POMDP), which is computationally intractable in general \cite{Bernstein2002}. Under prolonged signaling silences, the lack of a globally shared state makes direct centralized gradient-based coordination difficult under partial observability. These challenges necessitate a decentralized learning approach, motivating the proposed EPIC framework. By leveraging federated reinforcement learning, EPIC decomposes the global objective into localized semantic policies, utilizing the $\mathcal{T}_{\mathrm{F-STI}}$ operator to maintain coordination consistency.

Under such extreme connectivity constraints, distributed digital twins suffer severe perception degradation. While the AoII precisely captures state discrepancies, its direct calculation requires real-time ground truth, making it inaccessible during signaling silences. Therefore, we introduce the VoI as an operational surrogate to proactively quantify error growth.

\begin{definition}[Value of Information]
While the ultimate mission objective is to maximize the global WCE score, assessing the necessity of communication requires quantifying the risk of performance degradation. Since the WCE score and the SGI act as complementary measures (as established in \eqref{eq:wce_final}), the VoI assessed by agent $i$ concerning neighbor $j$ is evaluated based on the inferred coverage deficit:
\begin{equation}\label{eq:voi_def}
    \mathrm{VoI}_{i,j}(t) = \hat{\Gamma}_{i,m_j}(t) \cdot \Delta_{i,j}(t),
\end{equation}
where $\hat{\Gamma}_{i,m_j}(t)$ denotes the SGI of grid $m_j$ (agent $j$'s inferred location) maintained in agent $i$'s local world model, explicitly representing the accumulated service deficit and information staleness relevant to synchronization. The temporal staleness is given by $\Delta_{i,j}(t)=\min\{t-U_{i,j}(t),T_{\mathrm{up}}\}$, where $U_{i,j}(t)$ denotes the timestamp of the last successful update from agent $j$.
\end{definition}

The resulting VoI is an operational synchronization-urgency score: high values indicate that a neighbor is associated with a severe service deficit while its state information is increasingly stale. Consequently, crossing a predefined VoI threshold triggers the federated calibration mechanism to rectify this inference drift, thereby prioritizing federated recalibration when accumulated inference drift becomes increasingly relevant to collective task performance.

\section{The EPIC Framework: Embodied Proactive Coordination with Federated Calibration}
\label{sec:EPIC_framework}

\subsection{Hierarchical Architecture and System Workflow}
\label{sec:arch_overview}

Here, ``embodied'' refers to the onboard and agent-local grounding of the semantic representation in physical state, kinematics, and task context, rather than to an embodied-learning paradigm.
As illustrated in Fig.~\ref{fig:EPIC_architecture}, EPIC operates as a semantic-aware hierarchical closed-loop system. The framework partitions the coordination challenge into three functional layers to bridge the gap between low-level kinetic sensing and high-level semantic alignment:
\begin{itemize}
    \item \textit{Layer 1: Kinetic Perception \& Peer Inference}: Residing onboard the UAV, this layer processes the 14-D observation vector $\mathbf{o}_{i,t}$ and uses the $\mathcal{T}_{\mathrm{F-STI}}(\mathbf{s}_{j,t-\Delta_{i,j}(t)}, \alpha, \Delta_{i,j}(t))$ operator to reconstruct unavailable peer states. This enables the agent to extrapolate peer trajectories and maintain local world-model continuity during communication voids, with $0 < \Delta_{i,j}(t) \le T_{\mathrm{up}}$. 
    \item \textit{Layer 2: Federated Alignment \& VoI-Triggered Sync}: This infrastructure layer uses an edge base station to aggregate local parameters $\phi_i$ via an active-only scheme. A VoI threshold triggers uplink updates when stale peer information is associated with severe service deficits, while the global model $\Phi_G$ provides a shared reference for mission-aware coordination.   
    \item \textit{Layer 3: Proactive Execution \& Vector Navigation}: The agent synthesizes the calibrated semantic world model $\mathcal{V}_i(t)$ by fusing locally inferred task semantics with the federated global model through the dynamic calibration gain $\omega_t$. The resulting task-oriented digital twin is consumed directly by the stochastic policy $\pi(\mu,\sigma)$ to generate vector navigation commands $\mathbf{a}_{i,t}\in\mathbb{R}^2$, enabling autonomous resilient reconfiguration maneuvers to alleviate service gaps.
\end{itemize}

\subsection{Embodied Feature Representation and Multi-Modal Execution}
\label{sec:embodied_features_fsti}

\begin{figure*}[!t]    
    \centering
    \includegraphics[width=.8\textwidth]{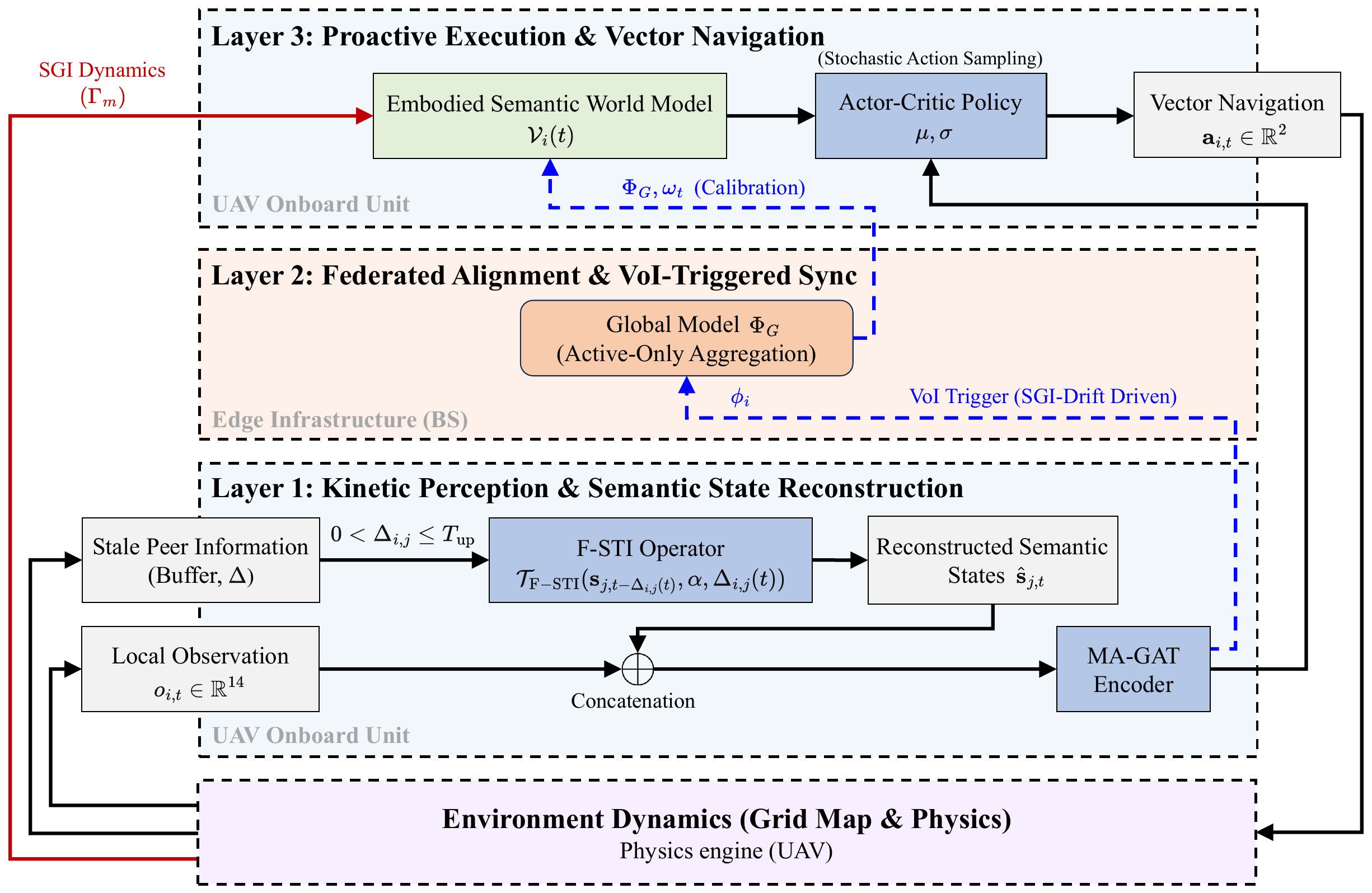}
    \caption{Hierarchical data flow and closed-loop execution architecture of the EPIC framework. The framework bridges low-level physical dynamics and high-level mission control across three functional layers: (1) Layer 1 handles kinetic perception, peer tracking, and Value of Information (VoI) evaluation; (2) Layer 2 governs the asynchronous, active-only federated alignment and semantic world-model calibration; and (3) Layer 3 executes vector-guided navigation and distributed recovery maneuvers to suppress service gaps under intermittent signaling. Solid black arrows denote the real-time onboard open-loop inference and control pipeline; dashed blue arrows represent the opportunistic closed-loop federated calibration pathway triggered by VoI; and the solid red arrow denotes the macro-level mission demand dynamics driven by the SGI, $\Gamma_m$, from the environment.}
    \label{fig:EPIC_architecture}
    \vspace{-1em}
\end{figure*}

As illustrated in Fig.~\ref{fig:EPIC_architecture}, the 14-dimensional observation vector $\mathbf{o}_{i,t} \in \mathbb{R}^{14}$ is partitioned into four functional groups: (1) \textit{Self-Kinematics} (4-D), comprising normalized 2D position $\mathbf{p}_{\mathrm{norm}}$ and velocity $\mathbf{v}_{\mathrm{norm}}$; (2) \textit{Boundary Awareness} (4-D), containing clearances to the four airspace boundaries; (3) \textit{Radio and Guidance} (3-D), encapsulating normalized RSRP $P_{i,k}(t)$, relative distance to the highest-deficit hotspot $d_{i,\mathrm{best}}$, and collision-danger signal $s_{\mathrm{danger}}$; and (4) \textit{Task Topology} (3-D), including nearest-neighbor relative coordinates $\mathbf{t}_{\mathrm{topo}}$, target direction vector $\mathbf{d}_{\mathrm{dir}}$, and normalized mission progress $t_{\mathrm{norm}}$.

When a peer agent $u_j$ becomes unreachable due to signaling silences ($\Delta_{i,j}(t) > 0$), agent $u_i$ invokes the F-STI operator, $\mathcal{T}_{\mathrm{F-STI}}$, to extrapolate the missing state features. Unlike zero-order hold methods that assume static positions, $\mathcal{T}_{\mathrm{F-STI}}$ leverages inherent UAV mobility constraints to maintain trajectory continuity. The inferred state $\hat{\mathbf{s}}_{j,t} = \mathcal{T}_{\mathrm{F-STI}}(\mathbf{s}_{j,t-\Delta_{i,j}(t)}, \alpha, \Delta_{i,j}(t))$ is recursively extrapolated from the last verified observation:
\begin{align}
    \hat{\mathbf{p}}_{j,t} &= \mathrm{clip} \left( \hat{\mathbf{p}}_{j,t-1} + \hat{\mathbf{v}}_{j,t-1} \delta t, \, 0, \, L \right), \label{eq:sti_pos_rec} \\
    \hat{\mathbf{v}}_{j,t} &= \hat{\mathbf{v}}_{j,t-1} (1 - \alpha), \label{eq:sti_vel_rec}
\end{align}
where $\alpha \in [0, 1]$ serves as the physical damping coefficient dictating the STI decay rate. By mapping the elapsed AoI $\Delta_{i,j}(t)$ to these inertia dynamics, the operator effectively suppresses coordination drift and bounds the finite-horizon peer-state reconstruction error within the ESWM $\mathcal{V}_i(t)$. 

To fuse these inferred peer states $\hat{\mathbf{s}}_{j,t}$ and the 14-D embodied vector $\mathbf{o}_{i,t}$ with grid-based semantics, the actor network employs a hierarchical multi-modal architecture. Initially, a 3-channel CNN processes $30 \times 30$ grid observations: a \textit{Weight Channel} encoding the static mission demand $w_m(t)$; an \textit{SGI Channel} tracking the SGI $\Gamma_m(t)$, which undergoes progressive recovery at rate $\delta_{\mathrm{rec}}$; and an \textit{Occupancy Channel} reflecting the spatial distribution of local and inferred neighbor positions to facilitate collision avoidance. The flattened CNN features are then concatenated with $\mathbf{o}_{i,t}$ and processed through a Mission-Aware Graph Attention (MA-GAT) layer, where attention coefficients are scaled by $1/\sqrt{d_{\mathrm{hidden}}}$ to maintain numerical stability. Finally, a stochastic policy head outputs the mean $\mu$ and standard deviation $\sigma$ of a Normal distribution, with a $\operatorname{softplus}$ activation applied to $\sigma$ to ensure positive standard deviations for continuous vector navigation.

\subsection{Federated Alignment and Active-Only Aggregation}
\label{sec:federated_alignment}

To mitigate the inference drift inherent in decentralized execution, EPIC incorporates a federated alignment layer that periodically synchronizes distributed world models. 
To balance global semantic consistency and stringent uplink constraints, we employ an \textit{Asymmetric Synchronization Protocol}. Specifically, each autonomous agent performs a downlink semantic recalibration from the edge global model $\Phi_G$ every $T_{\mathrm{cal}} = 100$ time steps. This resets the downlink staleness counter $\Delta_{i,\mathrm{BS}}(t)$, maintaining alignment among the distributed task-oriented digital representations even when uplink opportunities are scarce. Conversely, uplink aggregation is executed at a sparser interval of $T_{\mathrm{agg}} = 300$. This 1:3 synchronization ratio reduces the nominal scheduled uplink frequency by 66.7\% relative to a symmetric 100-step schedule, excluding additional event-triggered uploads.

Beyond scheduled intervals, EPIC employs a hybrid, event-driven mechanism to maintain semantic integrity. An active agent $u_i$ initiates an uplink transmission of its local encoder parameters $\phi_i$ to the edge base station under two conditions: either the scheduled aggregation window is reached ($t \pmod{T_{\mathrm{agg}}} = 0$) or its maximum local Value of Information ($\max_j \mathrm{VoI}_{i,j}(t)$) exceeds a predefined calibration threshold $\delta_{\mathrm{th}}$, conditioned on the physical availability of the uplink connection ($\mathrm{SINR}_{i,\mathrm{BS}}(t) \ge \gamma_{\mathrm{th}}$). Rather than physically restoring an unavailable wireless link, the VoI trigger serves as an opportunistic safety valve: it prioritizes an emergency model upload as soon as link connectivity is re-established, mitigating accumulated inference drift before internal uncertainty destabilizes swarm coordination. 
Upon receiving these updates, the base station computes the aggregated global model $\Phi_G$. To prevent failure pollution from stale parameters of failed or deactivated UAVs, we implement an \textit{Active-Only Aggregation} scheme: 
\begin{equation}\label{eq:fl_agg}
    \Phi_{G} = \frac{\sum_{i=1}^{N} \mathbb{1} \left\{ i \in \mathcal{U}_{\mathrm{act}} \right\} \cdot \phi_i}{\sum_{i=1}^{N} \mathbb{1} \left\{ i \in \mathcal{U}_{\mathrm{act}} \right\}},
\end{equation}
where $N$ is the initial swarm size, and $\phi_i$ denotes the parameter subset of agent $i$ governing the shared spatial task-demand encoder (i.e., CNN layers). During local PPO training, these encoder parameters are updated jointly with the local actor through backpropagation of the PPO policy objective. For active agents that do not trigger an upload at the current decision epoch, the edge server aggregates using their latest received and stored parameter snapshots, while the localized policy heads $\pi_{\theta_i}$ remain agent-specific and the centralized critic $V_\psi$ is excluded from federated aggregation.

Once the base station broadcasts the aggregated $\Phi_G$, each agent $u_i$ fuses this global reference with its locally trained encoder parameters $\phi_i$ to synthesize the spatial task-demand component of the ESWM, $\mathcal{V}_i(t)$. This synthesis utilizes a decoding mapping $\Psi(\cdot)$ that maps neural feature parameters into a spatial task-demand distribution $\mathcal{V}_i(t) \in \mathbb{R}^M$, consumed directly as an actionable digital twin representation for navigation. To account for unreliable wireless channels, the fusion is governed by a dynamic calibration gain $\omega_t \in [0, 1]$:
\begin{equation}\label{eq:calibration_gain}
    \omega_t = 1 - \exp\left( -\lambda_{\mathrm{cal}} \cdot \Delta_{i, \mathrm{BS}}(t) \right),
\end{equation}
where $\Delta_{i, \mathrm{BS}}(t)$ is the downlink AoI, measuring the elapsed time since the last successful base station update, and $\lambda_{\mathrm{cal}} > 0$ dictates the exponential decay rate of local trust. The task-demand distribution is synthesized via a convex combination:
\begin{equation}\label{eq:belief_fusion}
    \mathcal{V}_i(t) = (1 - \omega_t) \Psi(\phi_i) + \omega_t \Psi(\Phi_{G}).
\end{equation}
Here, the federated global model $\Phi_G$ is interpreted as a slowly varying population-level prior rather than a temporally fresh physical-state estimate, providing a stable conservative anchor when local observations become stale. Upon a successful calibration event, the agent computes the dynamic calibration gain $\omega_t$ using the accumulated downlink staleness $\Delta_{i,\mathrm{BS}}(t)$ elapsed prior to reception, synthesizes the world model via \eqref{eq:belief_fusion}, and subsequently resets the downlink staleness counter $\Delta_{i,\mathrm{BS}}(t) \leftarrow 0$. When the elapsed downlink interval before calibration is short ($\Delta_{i, \mathrm{BS}} \ge 1$), $\omega_t$ is bounded away from zero by $\omega_{\min} = 1 - \exp(-\lambda_{\mathrm{cal}}) > 0$, balancing local sensing with global guidance. Conversely, as $\Delta_{i,\mathrm{BS}}$ grows during prolonged silences, $\omega_t$ approaches 1 exponentially, progressively shifting the fusion weight toward the federated global anchor $\Psi(\Phi_G)$.

It is important to clarify the state of the global model $\Phi_G$ on the edge server. Unlike conventional synchronous Federated Learning that waits for a fixed window, our server acts as an \textit{asynchronous accumulator}. While the coarse-grained synchronization occurs every $T_{\mathrm{agg}}$ steps, the global model is continuously updated whenever any individual agent triggers an opportunistic VoI-based upload. Consequently, the periodic $T_{\mathrm{cal}}$ downloads performed by agents are not redundant; they allow individuals to retrieve fresh semantic insights contributed by peers who encountered high-priority mission events between the major $T_{\mathrm{agg}}$ intervals. This mechanism is intended to maintain a consistent and evolving collective representation even under sparse global updates.

\subsection{Federated Reinforcement Learning and Mission-Aware Reward}
\label{subsec:frl_reward}
While federated aggregation synchronizes global knowledge, the directional stability of decentralized execution relies strictly on aligning local incentives. To align individual policy incentives with the collective cooperative coverage objective, we meticulously shape a composite reward function $R_{i,t}$ for each agent $u_i$. To decentralize the cooperative coverage objective, we employ a fair-share distribution mechanism that assigns each covered cell's priority proportionally across the active UAVs serving that cell. This produces a diminishing-return congestion incentive when multiple UAVs cover the same region. To further prioritize regions with severe service deficits, we introduce a priority multiplier $\rho(\Gamma_m(t))$ based on the SGI. The task reward is formulated as:
\begin{equation}\label{eq:reward_task}
    R_{i,t}^{\mathrm{task}} = \sum_{m \in \mathcal{G}} \frac{w_m(t) \cdot \rho(\Gamma_m(t))}{\max\{\sum_{j \in \mathcal{U}_{\mathrm{act}}} \chi_{j,m}(t), 1\}} \cdot \chi_{i,m}(t),
\end{equation}
where $\chi_{i,m}(t) \in \{0,1\}$ is the instantaneous physical coverage indicator. The priority multiplier dynamically amplifies the reward for alleviating long-standing gaps:
\begin{equation}\label{eq:priority_multiplier}
    \rho(\Gamma_m(t)) = 1 + \frac{\Gamma_m(t)}{\Gamma_{\mathrm{th}}},
\end{equation}
with $\Gamma_{\mathrm{th}}$ being a normalizing threshold. 
Crucially, this fair-share allocation directly models the diminishing marginal return of multiple UAVs crowding into the same service zone. Rather than directly optimizing the non-linear WCE metric, this resource-sharing structure establishes a stage-game congestion utility whose marginal variations align with a Rosenthal-type harmonic potential function, as proven in Section~\ref{sec:theoretical_analysis}.

Beyond spatial coverage, the kinetic guidance reward $R_{i,t}^{\mathrm{nav}}$ aligns physical locomotion with semantic inference during signaling silences:
\begin{equation}\label{eq:reward_nav}
    R_{i,t}^{\mathrm{nav}} = \lambda_{\mathrm{nav}} \left( \mathbf{v}_{i,t} \cdot \mathbf{d}_{i, \mathrm{best}} \right),
\end{equation}
where $\mathbf{d}_{i,\mathrm{best}}$ is the unit vector pointing toward the highest-deficit hotspot identified by $\mathcal{V}_i(t)$. By regularizing velocity alignment, this directional gradient promotes smooth inertial maneuvers and prevents coordination deadlocks during communication voids.

To make the global optimization tractable for decentralized execution, these components are aggregated into the local utility used by the FMARL policy, providing a tractable decentralized realization of the system-level coverage objective. The total reward $R_{i,t}$ guiding the FMARL policy updates is aggregated as:
\begin{equation}\label{eq:total_reward_new}
    R_{i,t} = \frac{1}{\zeta} \left( R_{i,t}^{\mathrm{task}} + R_{i,t}^{\mathrm{nav}} - \lambda_{\mathrm{col}} \sum_{j \neq i} \mathbb{I}_{\mathrm{collision}}(i,j) \right) - \eta_{\mathrm{drift}} \epsilon_{\mathrm{drift}},
\end{equation}
where $\zeta$ is a scaling factor, and $\lambda_{\mathrm{col}}$ is the penalty weight associated with the minimum-separation violation indicator $I_{\mathrm{collision}}$, where $I_{\mathrm{collision}}(i,j)=1$ when $\|\mathbf p_i-\mathbf p_j\|<d_{\mathrm{safe}}$. Furthermore, we incorporate an inference drift penalty $\epsilon_{\mathrm{drift}}$ weighted by $\eta_{\mathrm{drift}}$. 
Crucially, this term acts strictly as an internal cognitive regularizer rather than a maneuver-dependent constraint. Specifically, $\epsilon_{\mathrm{drift}}$ is evaluated solely from peer-state reconstruction residuals $\|\mathbf{s}_{j,t} - \hat{\mathbf{s}}_{j,t}\|$ for $j \neq i$, rendering the penalty mathematically independent of agent $i$'s immediate physical control action $\mathbf{a}_{i,t}$ at the current decision epoch. The resulting reward is a tractable surrogate for the system-level objective in $\mathcal{P}_0$ rather than an exact reformulation; the subsequent game-theoretic analysis therefore characterizes the adopted local utility structure rather than global optimality of $\mathcal{P}_0$.

\subsection{Resilient Coordination and Symmetry Breaking}
\label{subsec:self_healing_mechanism}
To characterize decentralized coordination toward stable configurations, we consider the action-dependent component of the composite reward in \eqref{eq:total_reward_new} as the stage utility. The inference-drift regularizer is state-dependent and independent of the agent's immediate action at the current decision epoch, and therefore does not affect unilateral utility differences. This alignment makes unilateral action improvements consistent with the state-conditional congestion structure. However, a critical challenge in such decentralized execution is the \textit{homogeneous symmetry trap}, where identical agents execute identical actions, leading to collision clusters or redundant coverage. EPIC mitigates this through stochastic action sampling, where agents sample actions from a learned Gaussian distribution during execution:
\begin{equation}
\label{eq:stochastic_policy}
\begin{split}
    \tilde{\mathbf a}_i(t)&\sim\mathcal{N}\!\left(\boldsymbol{\mu}_{\theta}(\mathbf{o}_{i,t}),\boldsymbol{\sigma}_{\theta}(\mathbf{o}_{i,t})\right),\\
    \mathbf a_i(t)&=\operatorname{clip}\left(\tilde{\mathbf a}_i(t),-1,1\right).
\end{split}
\end{equation}
where the mean $\boldsymbol{\mu}_{\theta}$ and standard deviation $\boldsymbol{\sigma}_{\theta}$ are parameterized by the actor network. The learned standard deviation $\boldsymbol{\sigma}_{\theta}$ provides stochastic exploration perturbations, facilitating spatial dispersion while maintaining mission-oriented navigation.

\begin{remark}[Resilient Reconfiguration]
    \label{remark:self_healing_emergence}
    Proactive role transition occurs when the gradient of the vector navigation reward $R_{i,t}^{\mathrm{nav}}$ overcomes the perceived congestion penalty generated by reconstructed peer positions. This interaction incentivizes agents to autonomously migrate away from overcrowded regions and fill unattended service voids, realizing decentralized load-balancing.
\end{remark}

\begin{remark}[Symmetry Breaking and Equilibrium Realization]
    \label{remark:symmetry_breaking_convergence}
    The stochastic execution in \eqref{eq:stochastic_policy} introduces exploratory perturbations that can help escape suboptimal symmetric configurations on the state-conditional coordination surface, facilitating transitions toward stable equilibrium configurations.
\end{remark}

\subsection{Algorithm Execution and Complexity Analysis}
\label{subsec:algorithm_complexity}

To separate the stage-game equilibrium characterization from intermittent physical communications, the EPIC operational pipeline is designed under the Centralized Training with Decentralized Execution (CTDE) paradigm, as delineated in Algorithms \ref{alg:EPIC_Training} and \ref{alg:EPIC_Execution}. During offline training (Algorithm \ref{alg:EPIC_Training}), a global critic $V_\psi$ stabilizes decentralized policy learning, while the drift error $\epsilon_{\mathrm{drift}}$ is explicitly penalized to ground the F-STI operator $\mathcal{T}_{\mathrm{F-STI}}$ in physical kinematics, thereby reducing the state-reconstruction discrepancy underlying AoII. In the online deployment phase (Algorithm \ref{alg:EPIC_Execution}), each UAV operates as a fully autonomous embodied agent. The F-STI module compensates for signaling silences by recursively applying the damping factor $\alpha$ to extrapolate peer trajectories. To ensure consistent mission alignment, the federated layer utilizes both a fixed asymmetric schedule ($T_{\mathrm{cal}}$, $T_{\mathrm{agg}}$) and the maximum local Value of Information ($\max_j \mathrm{VoI}_{i,j}(t)$) as dynamic triggers to recalibrate the ESWM $\mathcal{V}_i(t)$.

\begin{algorithm2e}[!t]
\footnotesize
%\color{blue}
\caption{Phase I: Centralized Training for EPIC}
\label{alg:EPIC_Training}
\SetKwInOut{Input}{Input}\SetKwInOut{Output}{Output}
\Input{Mission grids $\mathcal{G}$, Outage bound $T_{\mathrm{up}}$, Kinematic damping $\alpha$, Sync periods $\{T_{\mathrm{cal}}, T_{\mathrm{agg}}\}$, VoI threshold $\delta_{\mathrm{th}}$}
\Output{Optimized actor parameters $\{\theta_i\}$, critic $\psi$, and F-STI operator $\mathcal{T}_{\mathrm{F-STI}}$}
Initialize global critic $V_{\psi}$, decentralized actors $\{\pi_{\theta_i}\}$, and F-STI operator $\mathcal{T}_{\mathrm{F-STI}}$\;
\For{each training episode $ep$}{
    Reset environment; initialize AoI matrices $\Delta_{i,j} \leftarrow 0$, $\Delta_{i,\mathrm{BS}} \leftarrow 0$\;
    Sample stochastic node failures $\{t_{\mathrm{fail}}, \mathcal{U}_{\mathrm{fail}}\}$ based on failure rate $\eta_{\mathrm{fail}}$\;
    \For{each time step $t \in \{1, \dots, T_{\max}\}$}{
        \For{each active agent $u_i \in \mathcal{U}_{\mathrm{act}}$}{
            \tcp{Layer 1: Belief Maintenance via F-STI}
            \For{each peer $u_j \in \mathcal{U} \setminus \{u_i\}$}{
                \eIf{$\mathrm{SINR}_{i,j}(t) \ge \gamma_{\mathrm{th}}$}{
                    Synchronize verified state $\mathbf{s}_{j,t}$ and reset $\Delta_{i,j} \leftarrow 0$\;
                }{
                    Increment AoI $\Delta_{i,j} \leftarrow \min(\Delta_{i,j} + 1, T_{\mathrm{up}})$\;
                    Extrapolate $\hat{\mathbf{s}}_{j,t} \leftarrow \mathcal{T}_{\mathrm{F-STI}}(\mathbf{s}_{j,t-\Delta_{i,j}}, \alpha, \Delta_{i,j})$ via \eqref{eq:sti_pos_rec}--\eqref{eq:sti_vel_rec}\;
                }
            }
            Extract observation $\mathbf{o}_{i,t}$ using local world model $\mathcal{V}_i(t)$ and inferred peer states $\{\hat{\mathbf{s}}_{j,t}\}$\;
            Sample navigation action $\mathbf{a}_{i,t} \sim \pi_{\theta_i}(\mathbf{o}_{i,t})$\;
        }
        Execute joint actions; update SGI map $\Gamma_m(t)$ via \eqref{eq:SGI_evolution}\;
        Collect stage rewards $R_{i,t}$ in \eqref{eq:total_reward_new} and evaluate drift loss $\epsilon_{\mathrm{drift}}$\;
        Store transition tuples in MAPPO rollout buffer\;
        \If{PPO update interval reached}{
            Update critic $\psi$ and actors $\{\theta_i\}$ via clipped surrogate objective and drift penalty\;
        }
        \tcp{Layer 2: Two-Tier Asymmetric Federated Synchronization}
        \For{each agent $u_i \in \mathcal{U}_{\mathrm{act}}$}{
            \If{$(t \pmod{T_{\mathrm{agg}}} = 0 \text{ or } \max_j \mathrm{VoI}_{i,j}(t) > \delta_{\mathrm{th}}) \text{ and } \mathrm{SINR}_{i,\mathrm{BS}}(t) \ge \gamma_{\mathrm{th}}$}{
                Transmit local encoder parameters $\phi_i$ to Edge Server\;
            }
        }
        \If{Edge Server receives parameter updates}{
            Aggregate $\Phi_G \leftarrow \sum_{i \in \mathcal{U}_{\mathrm{act}}} \phi_i / |\mathcal{U}_{\mathrm{act}}|$ via \eqref{eq:fl_agg}\;
        }
        \For{each agent $u_i \in \mathcal{U}_{\mathrm{act}}$}{
            \eIf{$(t \pmod{T_{\mathrm{cal}}} = 0 \text{ or Server Broadcasted}) \text{ and } \mathrm{SINR}_{i,\mathrm{BS}}(t) \ge \gamma_{\mathrm{th}}$}{
                Download $\Phi_G$\;
                Compute $\omega_t \leftarrow 1 - \exp(-\lambda_{\mathrm{cal}} \Delta_{i,\mathrm{BS}})$ via \eqref{eq:calibration_gain}\;
                Synthesize $\mathcal{V}_i(t) \leftarrow (1-\omega_t)\Psi(\phi_i) + \omega_t\Psi(\Phi_G)$\;
                Reset downlink AoI $\Delta_{i,\mathrm{BS}} \leftarrow 0$\;
            }{
                Increment downlink AoI $\Delta_{i,\mathrm{BS}} \leftarrow \Delta_{i,\mathrm{BS}} + 1$\;
                Update calibration gain $\omega_t$ via \eqref{eq:calibration_gain}\;
            }
        }
    }
}
\end{algorithm2e}

\begin{algorithm2e}[!t]    
\footnotesize
%\color{blue}
\caption{Phase II: Decentralized Resilient Execution}
\label{alg:EPIC_Execution}
\SetKwInOut{Input}{Input}\SetKwInOut{Output}{Output}
\Input{Trained policy parameters $\{\theta_i, \mathcal{T}_{\mathrm{F-STI}}\}$, VoI threshold $\delta_{\mathrm{th}}$, Sync periods $\{T_{\mathrm{cal}}, T_{\mathrm{agg}}\}$}
\Output{Real-time continuous control actions $\mathbf{a}_{i,t}$}
\For{each mission time step $t = 1, 2, \dots$}{
    \For{each autonomous agent $u_i \in \mathcal{U}_{\mathrm{act}}$}{
        \tcp{Layer 1: Peer Tracking and Belief Inference}
        \For{each peer $u_j \in \mathcal{U} \setminus \{u_i\}$}{
            \eIf{$\mathrm{SINR}_{i,j}(t) \ge \gamma_{\mathrm{th}}$}{
                Receive state $\mathbf{s}_{j,t}$; reset tracking AoI $\Delta_{i,j}(t) \leftarrow 0$\;
            }{
                Increment tracking AoI $\Delta_{i,j}(t) \leftarrow \min(\Delta_{i,j}(t-1) + 1, T_{\mathrm{up}})$\;
                Predict $\hat{\mathbf{s}}_{j,t} \leftarrow \mathcal{T}_{\mathrm{F-STI}}(\mathbf{s}_{j,t-\Delta_{i,j}(t)}, \alpha, \Delta_{i,j}(t))$ with boundary clipping $[0, L]$\;
            }
        }
        Evaluate local risk metric $\mathrm{VoI}_{i,\max}(t) = \max_j (\hat{\Gamma}_{i,m_j}(t) \cdot \Delta_{i,j}(t))$\;

        \tcp{Layer 2: Opportunistic Federated Calibration}
        \If{$(\mathrm{VoI}_{i,\max}(t) > \delta_{\mathrm{th}} \text{ or } t \pmod{T_{\mathrm{agg}}} = 0) \text{ and } \mathrm{SINR}_{i,\mathrm{BS}}(t) \ge \gamma_{\mathrm{th}}$}{
            Transmit local encoder parameters $\phi_i$ to base station\;
        }
        \eIf{$(t \pmod{T_{\mathrm{cal}}} = 0 \text{ or Global Model Broadcasted}) \text{ and } \mathrm{SINR}_{i,\mathrm{BS}}(t) \ge \gamma_{\mathrm{th}}$}{
            Receive aggregated global parameters $\Phi_G$\;
            Compute dynamic gain $\omega_t \leftarrow 1 - \exp(-\lambda_{\mathrm{cal}} \Delta_{i,\mathrm{BS}}(t))$\;
            Synthesize world model $\mathcal{V}_i(t) \leftarrow (1-\omega_t)\Psi(\phi_i) + \omega_t \Psi(\Phi_G)$\;
            Reset downlink staleness $\Delta_{i,\mathrm{BS}}(t) \leftarrow 0$\;
        }{
            Increment downlink staleness $\Delta_{i,\mathrm{BS}}(t) \leftarrow \Delta_{i,\mathrm{BS}}(t-1) + 1$\;
            Update trust weight $\omega_t$ via \eqref{eq:calibration_gain}\;
        }

        \tcp{Layer 3: Vector Navigation and Locomotion}
        Formulate observation $\mathbf{o}_{i,t}$ using calibrated model $\mathcal{V}_i(t)$ and inferred states $\{\hat{\mathbf{s}}_{j,t}\}$\;
        Extract guidance direction $\mathbf{d}_{i,\mathrm{best}}$ toward highest unfulfilled deficit cell\;
        Sample and execute control action $\mathbf{a}_{i,t}$ according to \eqref{eq:stochastic_policy}\;
        Track proximity sensors to monitor $d_{\mathrm{safe}}$\;
    }
}
\end{algorithm2e}

To characterize the computational scalability of EPIC for larger LAE deployments, we evaluate its dominant onboard coordination overhead. First, for \textit{Spatial-Topological Encoding}, the Graph Attention (GAT) mechanism processes observations with a per-step complexity of $\mathcal{O}(N \cdot |\mathcal{N}_{\mathrm{max}}| \cdot d^2)$, where $d$ is the embedding dimension. Since the maximum neighborhood size satisfies $|\mathcal{N}_{\mathrm{max}}| \ll N$ in practice, this exhibits linear scalability relative to the swarm population. Second, the \textit{Kinetic Peer Inference} ($\mathcal{T}_{\mathrm{F-STI}}$) performs recursive extrapolation scaling at $\mathcal{O}(N \cdot N_{\mathrm{silence}})$, where $N_{\mathrm{silence}}$ is the number of unreachable peers. Even under a total signaling blackout, this bounds at $\mathcal{O}(N^2)$, which remains compatible with low-overhead embedded execution under the evaluated swarm sizes. Finally, for \textit{Federated Anchor}, the federated aggregation overhead scales with the number of participating agents and the shared encoder parameter size $|\phi|$, rather than directly with the environmental grid resolution. This decoupling ensures efficiency even in fine-grained mission spaces.

Aggregating these components, the dominant per-step onboard coordination complexity is characterized as:
\begin{equation}
    \mathcal{C}_{\mathrm{total}} = \mathcal{O}\left( N \cdot |\mathcal{N}_{\mathrm{max}}| \cdot d^2 + N \cdot N_{\mathrm{silence}} \cdot \chi_{\mathrm{F-STI}} + N \cdot \chi_{\pi} \right), 
\end{equation}
where $N_{\mathrm{silence}}\le N-1$ denotes the number of unreachable peers per agent, while $\chi_{\mathrm{F-STI}}$ and $\chi_\pi$ denote the per-inference computational costs of F-STI and the actor network, respectively. This complexity structure keeps the per-step algorithmic overhead bounded under the evaluated communication conditions, while the low-latency execution behavior is evaluated empirically in Section~\ref{sec:simulation_results}.

\section{Theoretical Analysis}
\label{sec:theoretical_analysis}

This section establishes the analytical foundation of the EPIC framework under intermittent connectivity. The communication model is explicitly scoped to the resulting link availability, while PHY-layer resource allocation and detailed channel dynamics are outside the scope of this work. Accordingly, rather than assuming idealized static convergence, we develop a sequential chain of results that connects open-loop physical inference to semantic decision robustness and, ultimately, equilibrium robustness under intermittent information. Specifically, the analysis proceeds as follows:
\begin{enumerate}
    \item \textit{Kinematic error bounding}: We establish a finite-horizon bound on the physical tracking error induced by F-STI during signaling voids.
    \item \textit{Federated semantic-model alignment}: We characterize the contraction of anchor-relative semantic-model disagreement around a shared federated semantic anchor and bound the residual discrepancy.
    \item \textit{Cooperative coverage and congestion interaction}: We define the cooperative coverage objective and formulate the resulting interaction as a state-conditional congestion game with a potential structure.
    \item \textit{Approximate equilibrium under information mismatch}: Finally, we bound the resulting utility perturbation and establish the corresponding $\epsilon$-Nash robustness.
\end{enumerate}

\subsection{Kinematic Error and Semantic Robustness}
\label{subsec:kinematic_semantic_stability}

During communication voids, each agent executes open-loop state reconstruction via the F-STI operator. We first upper-bound the physical tracking discrepancy over a finite silence interval of $k$ discrete steps, where $k \in \{0,\ldots,T_{\mathrm{up}}\}$, and subsequently map this spatial bound into semantic coverage classification boundaries.

\begin{lemma}[Finite-Horizon Kinematic Inference Bound]
    \label{lemma:drift}
    Assume the true UAV translational acceleration is upper-bounded by $\|\mathbf{a}_t\| \le a_{\max}$ during the signaling silence interval, and let the F-STI damping factor satisfy $0\le\alpha\le1$. Given the F-STI recursive velocity extrapolation $\hat{\mathbf{v}}_{t+k} = (1-\alpha)^k \hat{\mathbf{v}}_t$, the spatial inference error $\epsilon_k = \|\mathbf{p}_{t+k} - \hat{\mathbf{p}}_{t+k}\|$ after $k$ discrete steps of duration $\delta t$ is upper-bounded by
    \begin{equation}\label{eq:drift_bound}
        \epsilon_k \le \delta t \sum_{\ell=0}^{k-1}
        \left(
        \|\mathbf{v}_{t} - \hat{\mathbf{v}}_{t}\|
        + a_{\max} \ell \delta t
        + \|\hat{\mathbf{v}}_t\|
        \big(1 - (1-\alpha)^\ell \big)
        \right).
    \end{equation}
\end{lemma}

\begin{proof}
    The true velocity evolves as $\mathbf{v}_{t+k} = \mathbf{v}_t + \sum_{\ell=0}^{k-1} \mathbf{a}_{t+\ell} \delta t$. Applying the triangle inequality yields the step-wise velocity discrepancy:
    \begin{equation}
        \|\mathbf{v}_{t+\ell} - \hat{\mathbf{v}}_{t+\ell}\| \le \|\mathbf{v}_t - \hat{\mathbf{v}}_t\| + a_{\max} \ell \delta t + \|\hat{\mathbf{v}}_t\|(1 - (1-\alpha)^\ell).
    \end{equation}
    Using the discrete kinematic recursions within the operational region, $\mathbf{p}_{t+\ell+1} = \mathbf{p}_{t+\ell} + \mathbf{v}_{t+\ell}\delta t$ and $\hat{\mathbf{p}}_{t+\ell+1} = \hat{\mathbf{p}}_{t+\ell} + \hat{\mathbf{v}}_{t+\ell}\delta t$, the accumulated spatial discrepancy unrolls as:
    \begin{equation}
        \mathbf{p}_{t+k} - \hat{\mathbf{p}}_{t+k} = (\mathbf{p}_t - \hat{\mathbf{p}}_t) + \delta t \sum_{\ell=0}^{k-1} (\mathbf{v}_{t+\ell} - \hat{\mathbf{v}}_{t+\ell}).
    \end{equation}
    Since the position is verified at timestamp $t$ ($\mathbf{p}_t = \hat{\mathbf{p}}_t$), taking the Euclidean norm and applying the triangle inequality yields the bounded accumulation in \eqref{eq:drift_bound}.
\end{proof}

For comparison, persistent-velocity dead reckoning yields approximately $\mathcal{O}(k v_{\max}\delta t)$ position-error growth. We next characterize how the bounded error $\epsilon_k$ affects semantic coverage decisions.

\begin{proposition}[Semantic Misclassification Condition]\label{prop:semantic_error}
    Let the true binary coverage status of grid $m$ be $\chi_{i,m}(\mathbf{p}) = \mathbb{1}\{\|\mathbf{p} - \mathbf{g}_m\| \le R_{\mathrm{cov}}\}$. If the spatial inference error is upper-bounded by $\|\mathbf{p} - \hat{\mathbf{p}}\| \le \epsilon_k$ (derived in Lemma \ref{lemma:drift}), a semantic misclassification $\chi_{i,m}(\mathbf{p}) \neq \chi_{i,m}(\hat{\mathbf{p}})$ can occur only if the inferred distance to hotspot $m$ lies within the boundary uncertainty band:
    \begin{equation}\label{eq:semantic_condition}
        \left| \|\hat{\mathbf{p}} - \mathbf{g}_m\| - R_{\mathrm{cov}} \right| \le \epsilon_k.
    \end{equation}
\end{proposition}

\begin{proof}
    By the reverse triangle inequality, the physical distance satisfies:
    \begin{equation}
        \Big| \|\mathbf{p} - \mathbf{g}_m\| - \|\hat{\mathbf{p}} - \mathbf{g}_m\| \Big| \le \|\mathbf{p} - \hat{\mathbf{p}}\| \le \epsilon_k.
    \end{equation}
    If $\|\hat{\mathbf{p}} - \mathbf{g}_m\| > R_{\mathrm{cov}} + \epsilon_k$, then $\|\mathbf{p} - \mathbf{g}_m\| \ge \|\hat{\mathbf{p}} - \mathbf{g}_m\| - \epsilon_k > R_{\mathrm{cov}}$, deterministically guaranteeing $\chi_{i,m}(\mathbf{p}) = \chi_{i,m}(\hat{\mathbf{p}}) = 0$. Conversely, if $\|\hat{\mathbf{p}} - \mathbf{g}_m\| < R_{\mathrm{cov}} - \epsilon_k$, then $\|\mathbf{p} - \mathbf{g}_m\| \le \|\hat{\mathbf{p}} - \mathbf{g}_m\| + \epsilon_k < R_{\mathrm{cov}}$, deterministically guaranteeing $\chi_{i,m}(\mathbf{p}) = \chi_{i,m}(\hat{\mathbf{p}}) = 1$. Hence, a discrepancy $\chi_{i,m}(\mathbf{p}) \neq \chi_{i,m}(\hat{\mathbf{p}})$ can occur only if \eqref{eq:semantic_condition} holds.
\end{proof}

Thus, semantic ambiguity is confined to a radial band of width $2\epsilon_k$ around the coverage boundary $R_{\mathrm{cov}}$.

\subsection{Bounded Semantic-Model Disagreement Relative to Federated Anchor}
\label{subsec:federated_anchor_consistency}

During open-loop phases, agents update their local world models $\mathcal{V}_i(t)$ based on private, asynchronous observation sequences, causing distributed representations to drift. In the closed-loop phase, federated calibration realigns these models with respect to the broadcasted federated semantic anchor $\mathcal{V}_G(t) = \Psi(\Phi_G(t))$.

\begin{proposition}[Bounded Anchor-Relative Semantic-Model Disagreement]
\label{prop:anchor_disagreement}

Conditioning on a fixed active set $\mathcal U_{\mathrm{act}}$ with aggregation weights satisfying $\alpha_i\ge\alpha_{\min}>0$ and $\sum_{i\in\mathcal U_{\mathrm{act}}}\alpha_i=1$ over the analyzed calibration sequence, and on a federated semantic anchor reference $\mathcal V_G$, define the anchor-relative semantic-model disagreement as $D_t = \sum_{i \in \mathcal{U}_{\mathrm{act}}} \alpha_i \|\mathcal{V}_i(t)-\mathcal{V}_G\|^2$. Here, $\|\mathcal{V}_i(t)-\mathcal{V}_G\|$ quantifies disagreement directly in the decoded spatial task-demand space, and is therefore used as the semantic-space discrepancy rather than a neural-parameter-space distance. Moreover, for every active agent $i$,
\begin{equation}
\label{eq:individual_semantic_bound}
    \|\mathcal{V}_i(t)-\mathcal{V}_G\| \le \sqrt{\frac{D_t}{\alpha_i}} \le \sqrt{\frac{D_t}{\alpha_{\min}}}.
\end{equation} 
Let $\Delta_c \le T_{\mathrm{cal}}$ denote the interval between successive effective calibration events. Assume that:
\begin{enumerate}
    \item Over each open-loop interval $t \in [r\Delta_c,(r+1)\Delta_c)$, accumulated spatial extrapolation error and local updates lead to the pre-calibration bound
    \begin{equation}
    D_{(r+1)\Delta_c^-}
    \le
    D_{r\Delta_c}
    + c_1\Delta_c^2\epsilon_{\max}^2
    + c_2\varpi^2
    \end{equation}
    under maximum spatial error $\epsilon_{\max}$ and observation noise variance $\varpi^2$;

    \item At each calibration epoch $t_r=(r+1)\Delta_c$, active agents calibrate their models via the convex combination
    \begin{equation}
    \mathcal{V}_i^+
    =
    (1-\omega_{i,t_r})\mathcal{V}_i
    +\omega_{i,t_r}\mathcal{V}_G(t_r).
    \end{equation}
    During effective calibration events, participating agents satisfy $\omega_{i,t_r}\ge\omega_{\min}>0$. For the contraction analysis below, the federated semantic anchor reference is conditioned to remain fixed over the analyzed calibration sequence.
\end{enumerate}
Then, under this fixed-anchor analysis, each calibration step acts as a strict contraction on the anchor deviation with contraction factor $q=(1-\omega_{\min})^2\in[0,1)$. The resulting anchor-relative disagreement satisfies
\begin{equation}
    \label{eq:disagreement_bound}
    \limsup_{r\to\infty} D_{r\Delta_c}
    \le
    \frac{q}{1-q}
    \left(
    c_1\Delta_c^2\epsilon_{\max}^2+c_2\varpi^2
    \right).
\end{equation}
\end{proposition}

\begin{proof}
    By the first assumption, the pre-calibration disagreement satisfies
    \begin{equation}
        D_{(r+1)\Delta_c^-}
        \le
        D_{r\Delta_c}
        + c_1\Delta_c^2\epsilon_{\max}^2
        + c_2\varpi^2.
    \end{equation}
    Under the fixed-anchor condition and $\omega_{i,t_r}\ge\omega_{\min}$,
    \begin{equation}
        \|\mathcal{V}_i^+ - \mathcal{V}_G\|^2 = (1 - \omega_{i,t_r})^2 \|\mathcal{V}_i - \mathcal{V}_G\|^2 \le (1 - \omega_{\min})^2 \|\mathcal{V}_i - \mathcal{V}_G\|^2.
    \end{equation}
    With $q=(1-\omega_{\min})^2<1$, the post-calibration disagreement satisfies:
    \begin{align}
        D_{(r+1)\Delta_c} &= \sum_{i \in \mathcal{U}_{\mathrm{act}}} \alpha_i \|\mathcal{V}_i^+ - \mathcal{V}_G\|^2 \le q D_{(r+1)\Delta_c^-} \nonumber \\
        &\le q D_{r \Delta_c} + q(c_1 \Delta_c^2 \epsilon_{\max}^2 + c_2 \varpi^2).
    \end{align}
    Unrolling this linear recurrence relation over $R$ calibration rounds yields:
    \begin{equation}
        D_{R \Delta_c} \le q^R D_0 + q(c_1 \Delta_c^2 \epsilon_{\max}^2 + c_2 \varpi^2) \sum_{j=0}^{R-1} q^j.
    \end{equation}
    Taking $\limsup_{R \to \infty}$ yields the geometric series sum, concluding the steady-state bound.
\end{proof}

\begin{remark}[Protocol-Induced Contraction Constant]
    The condition $\omega_{\min} > 0$ arises naturally from the discrete-time execution protocol rather than an arbitrary mathematical requirement. Since scheduled calibration events occur across non-zero downlink intervals ($\Delta_{i,\mathrm{BS}} \ge 1$), evaluating \eqref{eq:calibration_gain} yields $\omega_{i,t_r} \ge 1 - \exp(-\lambda_{\mathrm{cal}}) \triangleq \omega_{\min} > 0$. Thus, the contraction coefficient $q=\exp(-2\lambda_{\mathrm{cal}})<1$ is determined by the trust-decay rate $\lambda_{\mathrm{cal}}$ under the adopted calibration protocol.
\end{remark}

Thus, the calibration mechanism provides a bounded anchor-relative
semantic-model disagreement under the fixed-anchor condition.

\subsection{Cooperative Coverage Objective and State-Conditional Congestion Game}
\label{subsec:self_healing_analysis}

To realize collective mission objectives in a decentralized manner, we separate the system-level cooperative coverage objective from the individual incentives induced by shared service demand:
\begin{itemize}
    \item \textbf{System-Level Cooperative Coverage Objective}: The instantaneous cooperative coverage objective specifies the network-wide service coverage over prioritized hotspots:
    \begin{equation}\label{eq:system_objective}
        W(\mathbf{a}; \mathbf{z}_t) = \sum_{m \in \mathcal{G}} q_m(t) \cdot \mathbb{1}\{n_m(\mathbf{a}) > 0\},
    \end{equation}
    where $q_m(t)=w_m(t)\rho(\Gamma_m(t))$ denotes the prioritized service demand, and $n_m(\mathbf{a})=\sum_{j\in\mathcal{U}_{\mathrm{act}}} \chi_{j,m}(\mathbf{a})$ is the occupancy of grid $m$.
    \item \textbf{Agent-Level Congestion Incentive}: To discourage redundant coverage, the cooperative objective is decentralized through a fair-share diminishing-return mechanism:
    \begin{align}
        U_i(\mathbf{a};\mathbf{z}_t)
        =
        \frac{1}{\zeta}\Big[
        &\sum_{m\in\mathcal{G}}
        \chi_{i,m}(\mathbf{a})
        \frac{q_m(t)}
        {\max\{n_m(\mathbf{a}),1\}}
        +R_{i,t}^{\mathrm{nav}}(\mathbf{a}_i;\mathbf{z}_t)
        \notag\\
        &-\lambda_{\mathrm{col}}
        \sum_{j\neq i}
        \mathbb{I}_{\mathrm{collision}}(i,j)
        \Big]. \label{eq:agent_utility_split}
    \end{align}
\end{itemize}

By construction,
\begin{equation}
    \sum_{i\in\mathcal{U}_{\mathrm{act}}}
    R_{i,t}^{\mathrm{task}}
    =
    W(\mathbf{a};\mathbf{z}_t),
\end{equation}
so the fair-share mechanism exactly decomposes the instantaneous coverage objective among active agents while inducing diminishing individual returns as the number of covering agents increases.

When an agent relocates from a redundant hotspot $m$ to an unattended hotspot $r$ satisfying $q_r>q_m/n_m$, its task utility strictly improves, providing a decentralized incentive for reallocating coverage toward underserved regions.

\begin{definition}[State-Conditional Congestion Game]
    \label{def:congestion_game}
    At decision epoch $t$, historical variables, including demand weights $w_m(t)$, SGI $\Gamma_m(t)$, and inferred peer positions, are treated as fixed exogenous state parameters $\mathbf{z}_t$. The interaction is modeled as a stage game $\mathcal{G}_t(\mathbf{z}_t) = \langle \mathcal{U}_{\mathrm{act}}, \{\mathcal{A}_i\}, \{U_i(\cdot; \mathbf{z}_t)\} \rangle$. For a fixed $\mathbf{z}_t$, navigation guidance $\mathbf{d}_{i,\mathrm{best}}$ is exogenous, ensuring $R_{i,t}^{\mathrm{nav}}(\mathbf{a}_i; \mathbf{z}_t)$ depends strictly on agent $i$'s own action.
\end{definition}

The congestion component arises from shared hotspot occupancy, while the navigation term is private-action dependent and the collision term induces a symmetric pairwise interaction.

\begin{theorem}[Equilibrium Characterization of the State-Conditional Congestion Game]
    \label{thm:potential_congestion_game}
    For every fixed state $\mathbf{z}_t$, the stage game $\mathcal{G}_t(\mathbf{z}_t)$ admits the following Rosenthal-type exact potential function:
    \begin{equation}\label{eq:exact_potential_func}
        \begin{split}
            \tilde{\Phi}_t(\mathbf{a}) = \frac{1}{\zeta} \Bigg[ &\sum_{m \in \mathcal{G}} q_m(t) H_{n_m(\mathbf{a})} + \sum_{i \in \mathcal{U}_{\mathrm{act}}} R_{i,t}^{\mathrm{nav}}(\mathbf{a}_i; \mathbf{z}_t) \\
            & - \frac{1}{2} \sum_{i \in \mathcal{U}_{\mathrm{act}}} \sum_{j \neq i} \lambda_{\mathrm{col}} \mathbb{I}_{\mathrm{collision}}(i,j) \Bigg],
        \end{split}
    \end{equation}
    where $H_n = \sum_{k=1}^n \frac{1}{k}$ denotes the harmonic number ($H_0 \triangleq 0$). For any unilateral deviation $a_i \to a_i'$, we have:
    \begin{equation}
        U_i(a_i', a_{-i}; \mathbf{z}_t) - U_i(a_i, a_{-i}; \mathbf{z}_t) = \tilde{\Phi}_t(a_i', a_{-i}) - \tilde{\Phi}_t(a_i, a_{-i}).
    \end{equation}
\end{theorem}

\begin{proof}
    Consider a unilateral action deviation by agent $i$ from $a_i$ to $a_i'$. Let $\mathcal{M}_i^- = \{m \in \mathcal{G} : \chi_{i,m}(a_i)=1, \chi_{i,m}(a_i')=0\}$ be abandoned cells, and $\mathcal{M}_i^+ = \{m \in \mathcal{G} : \chi_{i,m}(a_i)=0, \chi_{i,m}(a_i')=1\}$ be newly covered cells. The change in agent $i$'s task utility is
    \begin{equation}
        \Delta U_i^{\mathrm{task}}
        = \sum_{m \in \mathcal{M}_i^+} \frac{q_m(t)}{n_m + 1}
        - \sum_{m \in \mathcal{M}_i^-} \frac{q_m(t)}{n_m}.
    \end{equation}
    Concurrently, the variation in the harmonic component of the potential is
    \begin{equation}
        \begin{split}
            \Delta \tilde{\Phi}_t^{\mathrm{task}}
            &= \sum_{m \in \mathcal{M}_i^+} q_m(t)
            \bigl(H_{n_m+1} - H_{n_m}\bigr) \\
            &\quad + \sum_{m \in \mathcal{M}_i^-} q_m(t)
            \bigl(H_{n_m-1} - H_{n_m}\bigr) \\
            &= \sum_{m \in \mathcal{M}_i^+} \frac{q_m(t)}{n_m + 1}
            - \sum_{m \in \mathcal{M}_i^-} \frac{q_m(t)}{n_m}
            = \Delta U_i^{\mathrm{task}}.
        \end{split}
    \end{equation}
    The navigation term $R_{i,t}^{\mathrm{nav}}(\mathbf{a}_i;\mathbf{z}_t)$ depends only on agent $i$'s own action, so its unilateral variation is directly captured by the corresponding term in the potential. Likewise, the symmetric pairwise collision term is captured by the $\frac{1}{2}$ scaling in the potential to avoid double counting. Since utility and potential share the same positive normalization factor $\zeta$, the complete unilateral variation satisfies
    \begin{equation}
        \Delta U_i = \Delta \tilde{\Phi}_t.
    \end{equation}
\end{proof}

\begin{remark}[Equilibrium Existence under Finite Actions]
\label{rmk:finite_convergence}
    Under a canonical discretization of the operational action space $\mathcal{A}_i$ (or over finite navigational motion primitives), the stage game satisfies the finite improvement property (FIP) because $\tilde{\Phi}_t$ is an exact potential function. Every asynchronous strict better-response step strictly increases $\tilde{\Phi}_t$, precluding improvement cycles and guaranteeing the existence of at least one Pure-Strategy Nash Equilibrium (PNE). Under the finite-action condition, every PNE is a unilateral local maximum of the potential. This characterization does not imply that every PNE maximizes the system-level objective $W$ or is Pareto-optimal.
\end{remark}

\subsection{Approximate Equilibrium under Intermittent Connectivity}
\label{subsec:approximate_equilibrium}

Theorem~\ref{thm:potential_congestion_game} provides the complete-information equilibrium reference. Under intermittent connectivity, agents act on perceived states $\hat{\mathbf{z}}_{i,t}$ reconstructed from F-STI inference and calibrated world models. We next bound the induced utility perturbation.

Let the composite state uncertainty over outage interval $T_{\mathrm{up}}$ be bounded by
\begin{equation}
\varepsilon_z(T_{\mathrm{up}})
=
c_p \epsilon_{T_{\mathrm{up}}}
+c_v\sqrt{D_{T_{\mathrm{up}}}}
+c_\varpi\varpi,
\end{equation}
where $\epsilon_{T_{\mathrm{up}}}$ is bounded by Lemma~\ref{lemma:drift}, $D_{T_{\mathrm{up}}}$ is the aggregate anchor-relative semantic-model discrepancy in the decoded spatial task-demand space, and $c_v$ includes the factor $1/\sqrt{\alpha_{\min}}$ and the associated semantic-to-state scaling.

Proposition~\ref{prop:semantic_error} confines coverage-indicator mismatch to the $2\epsilon_k$-radial boundary band. We impose a boundary margin $\mu_\partial>0$ such that $\epsilon_k<\mu_\partial$ and admissible perturbations remain away from coverage and safety boundaries, together with a positive task-direction margin ensuring a unique highest-deficit hotspot. We further require a positive distance from that hotspot, so that the normalized guidance direction remains locally Lipschitz on the admissible region.

On the resulting compact operational region, assuming the neural decoder $\Psi(\cdot)$ is locally Lipschitz continuous \cite{Deligkas2020Lipschitz}, the stage utility admits a finite Lipschitz modulus $\bar{L}_U<\infty$. Consequently,
\begin{equation}\label{eq:lipschitz_bridge}
    \sup_{\mathbf{a} \in \mathcal{A}}
    \left| U_i(\mathbf{a}; \mathbf{z}_t) - \hat{U}_i(\mathbf{a}) \right|
    \le \delta_U(T_{\mathrm{up}})
    \le \bar{L}_U \varepsilon_z(T_{\mathrm{up}}).
\end{equation}

\begin{corollary}[$\epsilon$-Nash Equilibrium under Bounded Estimation Error]\label{cor:epsilon_nash}
    Let $\hat{U}_i(\mathbf{a}) \triangleq U_i(\mathbf{a};\hat{\mathbf{z}}_{i,t})$ denote agent $i$'s perceived stage utility, satisfying \eqref{eq:lipschitz_bridge}. Let $\hat{\mathbf a}^*$ be a pure-strategy Nash equilibrium of the perceived stage game $\hat{\mathcal G}_t$. Then, $\hat{\mathbf a}^*$ is an $\epsilon_{\mathrm{NE}}$-approximate Nash equilibrium of the true stage game $\mathcal G_t$, where
    \begin{equation}
        \epsilon_{\mathrm{NE}} \le 2 \delta_U(T_{\mathrm{up}}) \le 2 \bar{L}_U \varepsilon_z(T_{\mathrm{up}}).
    \end{equation}
\end{corollary}

\begin{proof}
    By definition of the equilibrium $\hat{\mathbf{a}}^*$ in the perceived stage game $\hat{\mathcal{G}}_t$, $\hat{U}_i(\hat{\mathbf{a}}^*) \ge \hat{U}_i(a_i', \hat{\mathbf{a}}_{-i}^*)$ holds for any unilateral alternative $a_i' \in \mathcal{A}_i$. Invoking the uniform perturbation bound \eqref{eq:lipschitz_bridge}:
    \begin{align}
        U_i(\hat{\mathbf{a}}^*) &\ge \hat{U}_i(\hat{\mathbf{a}}^*) - \delta_U(T_{\mathrm{up}}) \nonumber \\
        &\ge \hat{U}_i(a_i', \hat{\mathbf{a}}_{-i}^*) - \delta_U(T_{\mathrm{up}}) \nonumber \\
        &\ge U_i(a_i', \hat{\mathbf{a}}_{-i}^*) - 2 \delta_U(T_{\mathrm{up}}).
    \end{align}
    Therefore, the maximum unilateral regret is bounded by $2\delta_U(T_{\mathrm{up}})$, which vanishes as $\varepsilon_z(T_{\mathrm{up}})\to0$.
\end{proof}

\section{Simulation and Results}
\label{sec:simulation_results}

\subsection{Simulation Setup and Performance Metrics}
\label{subsec:setup_and_metrics}
The comprehensive parameters are detailed in Table~\ref{tab:parameters}. We evaluate the EPIC framework in a representative LAE simulation scenario covering an $1,800 \times 1,800 \times 100\,\mathrm{m}^3$ mission airspace. The UAVs cruise at a regulated altitude of $H_{\mathrm{uav}} = 100$~m, with horizontal navigation executed in the 2-D plane while vertical elevation is accounted for in 3D RSRP calculation with probabilistic LoS and shadowing. The environment is discretized into a $30 \times 30$ grid ($\mathcal{L}_{\mathrm{grid}} = 60\,\mathrm{m}$) featuring $K=8$ dynamic demand hotspots governed by the DGMM. The swarm comprises $N=6$ UAVs, each operating with an effective service radius $R_{\mathrm{cov}}=150\,\mathrm{m}$, a maximum velocity of $15\,\mathrm{m/s}$, and a discrete control time step of $\delta t = 1.0\,\mathrm{s}$, i.e., one simulation step corresponds to one second of physical time.

To assess system resilience, the training follows a two-phase curriculum learning protocol. Phase 1 (Episodes 1--500) optimizes baseline coverage under ideal signaling. Phase 2 (Episodes 501--1,000) introduces extreme asynchronous signaling, with observation staleness up to $T_{\mathrm{up}}=15$ steps, and dynamic node failures. During Phase 2, up to 3 agents may be stochastically deactivated between steps 80 and 200 to simulate catastrophic disruptions.

The decentralized policies use the attention-enhanced MARL architecture described in Section~\ref{sec:embodied_features_fsti}. The F-STI kinematic decay is set to $\alpha=0.25$. Furthermore, the asymmetric federated synchronization intervals are configured at $T_{\mathrm{cal}}=100$ and $T_{\mathrm{agg}}=300$, corresponding to a nominal 66.7\% reduction in scheduled uplink frequency relative to symmetric synchronization.

\begin{table}[!t]
%\color{blue}
\centering
\footnotesize
\caption{Simulation Parameters and EPIC System Configurations}
\label{tab:parameters}
\setlength{\tabcolsep}{3pt}
\begin{tabular}{@{}p{1.7cm}p{3.7cm}p{3cm}@{}}
\toprule
\textbf{Category} & \textbf{Parameter} & \textbf{Value (Description)} \\ \midrule

\multirow{6}{1.8cm}{Environment \& Dynamics} 
& Mission Area ($L \times L$) & $1800 \times 1800\,\mathrm{m}^2$ \\
& Grid Resolution & $30 \times 30$ ($L_{\mathrm{grid}} = 60\,\mathrm{m}$) \\
& Dynamic Hotspots ($K$) & 8 (Up to 20 for scalability) \\
& Target Weight ($w_m(t)$) & [1.0, 5.0] (Priority) \\
& SGI Aging/Recovery & 1.0 / 15.0 (units/step) \\ 
& Task Cutoff ($w_{\mathrm{th}}$) & 0.25 (Active demand threshold) \\ \midrule

\multirow{7}{1.8cm}{UAV Swarm \& Kinematics} 
& Number of Agents ($N$) & 6 (Up to 30 for scalability) \\
& Service Radius ($R_{\mathrm{cov}}$) & $150\,\mathrm{m}$ \\
& Max Velocity ($v_{\max}$) & $15.0\,\mathrm{m/s}$ \\
& Action Scaling ($v_{\mathrm{gain}}$) & $15.0\,\mathrm{m/s}$ \\ 
& Control Time Step ($\delta t$) & $1.0\,\mathrm{s}$ \\ 
& Safety Distance ($d_{\mathrm{safe}}$) & $30\,\mathrm{m}$ \\ 
& Kinetic Damping ($\alpha$) & 0.25 (F-STI prior) \\ \midrule

\multirow{11}{1.8cm}{Federated \& Comm.} 
& Global Sync ($T_{\mathrm{agg}}$) & 300 steps \\
& Local Sync ($T_{\mathrm{cal}}$) & 100 steps \\
& Signaling Void ($T_{\mathrm{up}}$) & 15 / 30 steps (Swept to 50) \\
& Transmit Power ($P_{\mathrm{tx}}$) & 23 dBm \\
& Carrier Frequency ($f_c$) & 2.4 GHz \\
& A2A Shadowing Margin ($\xi_{\mathrm{A2A}}$) & 3.0 dB \\
& A2G LoS Parameters ($a, b$) & (4.88, 0.43) (Suburban) \\
& Env. Attenuation ($\eta_{\mathrm{LoS}}, \eta_{\mathrm{NLoS}}$) & (0.1, 21) dB \\
& SINR Threshold ($\gamma_{\mathrm{th}}$) & -10 dB \\
& Noise Floor ($N_0$) & -104 dBm \\
& VoI Trigger ($\delta_{\mathrm{th}}$) & 100.0 (SGI-based) \\ \midrule

\multirow{8}{1.8cm}{MARL \& Training} 
& WCE Exponent ($\alpha_w$) & 1.5 \\
& Navigation Weight ($\lambda_{\mathrm{nav}}$) & 1.0 \\
& Separation-Envelope Violation Penalty ($\lambda_{\mathrm{col}}$) & 10.0 \\
& Inference Drift Penalty ($\eta_{\mathrm{drift}}$) & 5.0 (Reward shaping) \\
& Reward Scale ($\zeta^{-1}$) & $1/50$ \\
& Learning Rate & $3 \times 10^{-4}$ (MAPPO) \\
& Discount Factor ($\gamma$) & 0.99 \\
& PPO Clip ($\epsilon$) & 0.2 \\ \bottomrule
\end{tabular}
\end{table}

To quantify the resilience and efficiency of this simulated setup under sparse signaling constraints, we evaluate three multi-dimensional metrics. First, the WCE score evaluates the aggregate instantaneous mission fulfillment. As previously formulated in \eqref{eq:wce_final}, this metric captures true operational capacity by aggregating non-linear physical coverage across the grid set $\mathcal{G}$. 
Specifically, the exponent for the WCE score is set to $\alpha_w = 1.5$. It is crucial to note that this exponent is used exclusively during evaluation to emphasize coverage differences at high-priority hotspots. It does not participate in the internal reward shaping of the FMARL agents. During training, the agents rely entirely on the linear SGI-based priority multiplier defined in \eqref{eq:priority_multiplier}. This separation ensures that the non-linear evaluation metric does not alter the reward structure used in the game-theoretic analysis.

Second, the Service Outage Score (SOS) characterizes service deficiency across mission-critical regions by averaging the accumulated SGI:
\begin{equation}\label{eq:sos}
    \mathrm{SOS}(t) = \frac{1}{|\mathcal{G}_{\mathrm{task}}|} \sum_{m \in \mathcal{G}_{\mathrm{task}}} \Gamma_m(t),
\end{equation}
where $\mathcal{G}_{\mathrm{task}} = \{ m \in \mathcal{G} \mid w_m(t) > w_{\mathrm{th}} \}$ denotes the active subset of high-priority task cells. A lower average SOS demonstrates that proactive patrolling effectively minimizes service gaps despite extreme information aging.

Finally, the \textit{Swarm Alignment Score} (SAS) evaluates the precision of proactive maneuvers. Serving as a functional proxy for state-reconstruction accuracy, SAS measures the spatial similarity between the estimated trajectory and the ground-truth displacement:
\begin{equation}\label{eq:sas}
    \mathrm{SAS}_t = \exp \left( - \frac{1}{\sigma_{\mathrm{sas}} |\mathcal{U}_{\mathrm{act}}|} \sum_{i \in \mathcal{U}_{\mathrm{act}}} \|\mathbf{p}_{i,t} - \hat{\mathbf{p}}_{i,t}\| \right),
\end{equation}
where $\mathbf{p}_{i,t}$ and $\hat{\mathbf{p}}_{i,t}$ are the ground-truth and inferred positions of agent $i$, respectively. The scaling factor is set to $\sigma_{\mathrm{sas}} = 150\,\mathrm{m}$ to align with the UAV service radius $R_{\mathrm{cov}}$, ensuring high sensitivity to inference drifts that exceed physical coverage capabilities. A higher SAS indicates more accurate spatial state reconstruction by the F-STI operator, corresponding to lower cumulative inference error \(\epsilon_{\mathrm{drift}}\) during communication voids.

We evaluate EPIC against the following baselines and ablated variants under intermittent connectivity:
\begin{itemize}
    \item \textbf{EPIC (Full):} The proposed complete FMARL framework integrating the F-STI semantic world model for proactive inference, supported by the VoI-triggered asynchronous federated synchronization.
    \item \textbf{EPIC w/o VoI:} An ablated variant that retains the F-STI operator but relies entirely on strict periodic synchronization without the VoI trigger. This evaluates the efficiency of the opportunistic signaling mechanism.
    \item \textbf{EPIC w/o F-STI (ZOH):} An inference-time ablation of EPIC that replaces the F-STI-based peer-state reconstruction with a Zero-Order Hold (ZOH) mechanism, while keeping the trained policy parameters and other framework components unchanged.
    \item \textbf{MAPPO-KF:} An independently trained standard MAPPO baseline~\cite{yu2022surprising} in which unavailable neighbor states are estimated using a \textit{Kalman filter} (KF)~\cite{BarShalom2002} under intermittent connectivity, trained under identical outage distributions, action spaces, and control step durations.
\end{itemize}

\subsection{Simulation Result Analysis}

\subsubsection{Training Stability and Resilience}

\begin{figure}[!t]
    \centering
    \begin{subfigure}[b]{0.49\columnwidth}
        \centering
        \includegraphics[width=\linewidth]{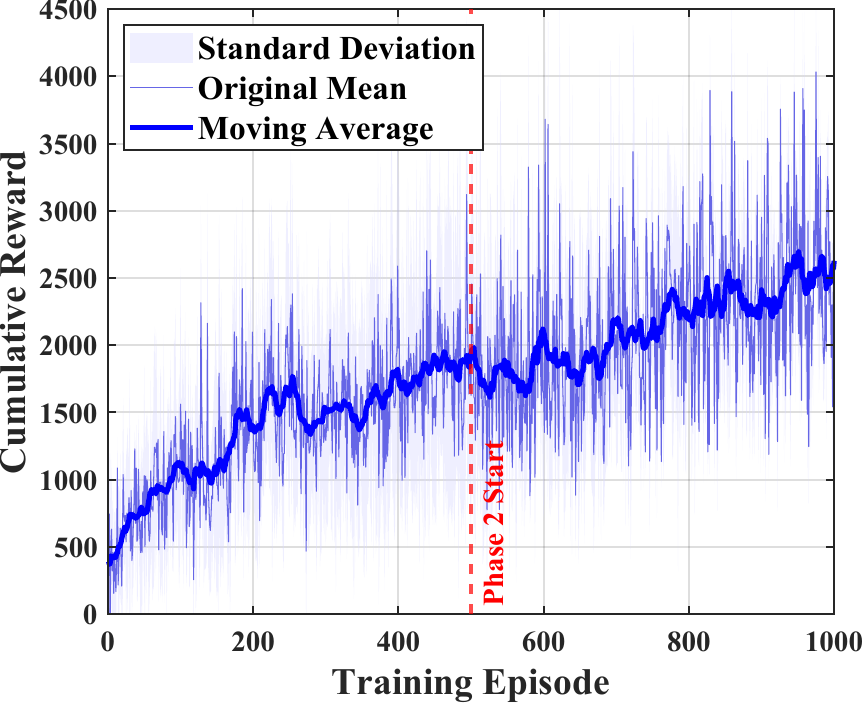}
        \caption{Total system reward}
        \label{fig:convergence_analysis:conv_reward}
    \end{subfigure}\hfill
    \begin{subfigure}[b]{0.49\columnwidth}
        \centering
        \includegraphics[width=\linewidth]{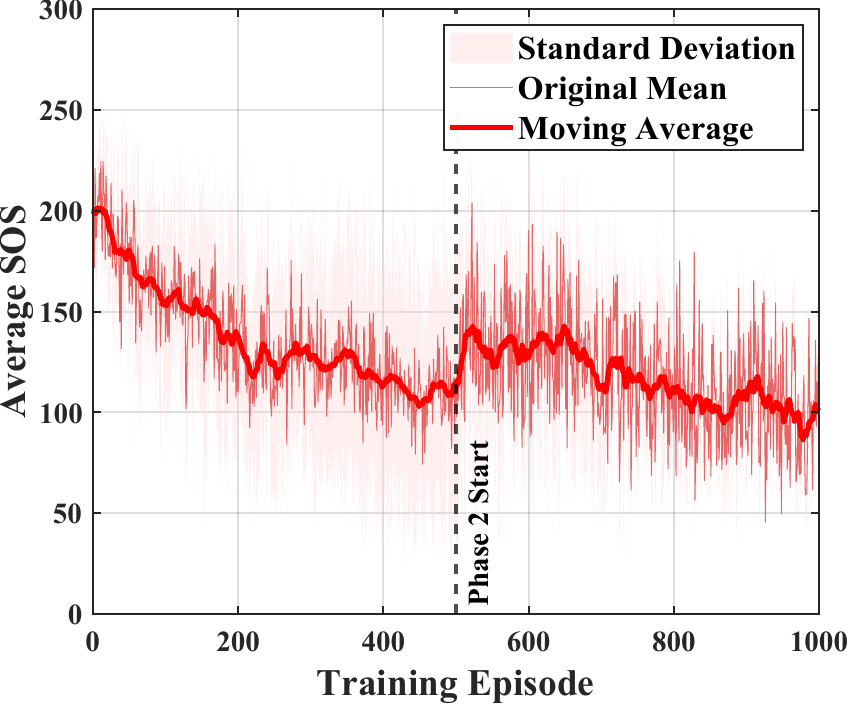}
        \caption{SOS convergence}
        \label{fig:convergence_analysis:conv_sos}
    \end{subfigure}    
    \caption{Training stabilization over five independent seeds. (a) Cumulative reward showing Phase 1 growth and rapid Phase 2 adaptation. (b) Mean SOS decay, demonstrating resilience under $T_{\mathrm{up}}=15$ staleness.}
    \label{fig:convergence_analysis}
\end{figure}

Fig.~\ref{fig:convergence_analysis} illustrates the training stabilization of the EPIC framework. During Phase 1 (Episodes 1--500), the average SOS exhibits a rapid monotonic decline, indicating the efficient capture of fundamental spatiotemporal demand patterns via the agents' embodied perception.

Introducing a 15-step signaling void at Episode 501 elevates tracking staleness and triggers a transient performance fluctuation. However, the swarm quickly recalibrates its coordination logic via the F-STI operator, reaching a stable performance regime by Episode 800. This resilience stems from progressive SOS recovery and $1/\zeta = 1/50$ reward scaling, which provide stable gradients for policy optimization during signaling silences.

%\color{blue}
Notably, while optimized under a nominal training outage of $T_{\mathrm{up}}=15$, subsequent evaluations adopt a doubled stress-test gap of $T_{\mathrm{up}}=30$ unless otherwise specified, with Figs. \ref{fig:mechanism_resilience} and \ref{fig:latency_staleness} further sweeping the signaling gap up to 50 steps to evaluate extreme latency limits.

\subsubsection{Generalization Analysis and Multi-Task Adaptability}
\label{ssbsec:cross_evaluation}
\begin{figure}[!t]
    \centering
    \begin{subfigure}[b]{0.49\columnwidth}
        \centering
        \includegraphics[width=\linewidth]{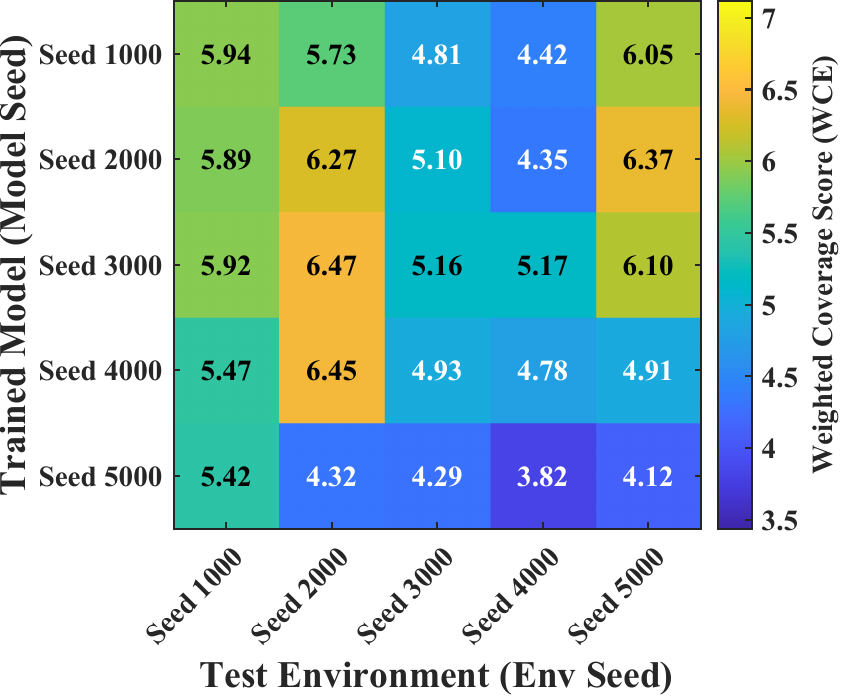}
        \caption{WCE Score}
        \label{fig:cross_evaluation:cross_eval_wce}
    \end{subfigure}\hfill
    \begin{subfigure}[b]{0.49\columnwidth}
        \centering
        \includegraphics[width=\linewidth]{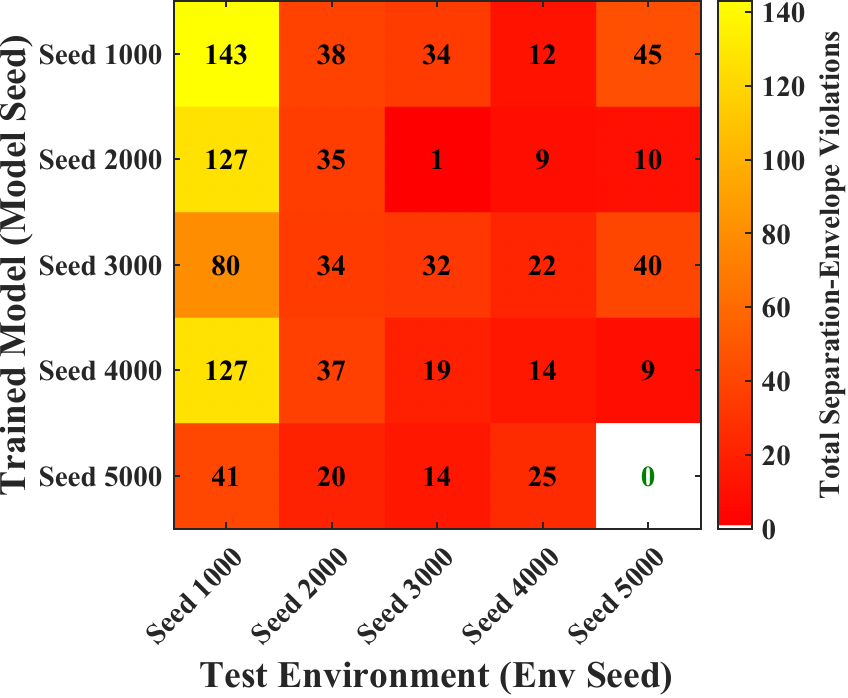}
        \caption{Separation-Envelope Violations}
        \label{fig:cross_evaluation:cross_eval_col}
    \end{subfigure}
    \caption{Cross-evaluation heatmaps across five independent seeds.
    (a) Off-diagonal WCE scores and 
    (b) cumulative separation-envelope
    violations ($d_{ij}<d_{\mathrm{safe}}=30$ m) over 600 steps.}
    \label{fig:cross_evaluation}
\end{figure}

To validate whether EPIC's spatiotemporal representations generalize beyond specific training instances, we conduct a cross-evaluation across five independently generated task environments (Seeds 1,000--5,000) over 600 operational steps. As shown in Fig.~\ref{fig:cross_evaluation:cross_eval_wce}, the off-diagonal WCE scores exhibit consistent cross-environment performance (3.82 to 6.47), indicating that the learned coordination policy generalizes across the independently generated deployment profiles. Fig.~\ref{fig:cross_evaluation:cross_eval_col} examines the corresponding spatial separation metrics under stochastic action sampling ($\mathbf{a}_i \sim \mathcal{N}(\boldsymbol{\mu}, \boldsymbol{\sigma}_\theta)$). In decentralized MARL with shared parameters, identical deterministic policies risk falling into a \textit{homogeneous symmetry trap}, wherein co-located agents compute identical greedy targets and produce persistent flight deadlocks. By leveraging the learned policy variance, EPIC introduces stochastic perturbations that can diversify multi-agent paths and mitigate persistent symmetric deadlocks. Across 9,000 pairwise evaluations per episode ($600~\text{steps} \times \binom{6}{2}$ pairs), separation breaches into the conservative $d_{\mathrm{safe}} = 30$~m exclusion envelope remain near zero in benign seeds, peaking at 143 instances ($<1.6\%$ event rate) only in Seed 1,000 where task hotspots heavily overlap. This indicates that stochastic execution effectively mitigates macro-level trajectory overlap without requiring centralized arbitration.

\subsubsection{Coordination Robustness and Safety Sensitivity}
\label{ssbsec:resilience_safty}
\begin{figure}[!t]
    \centering
    \begin{subfigure}[b]{0.485\columnwidth}
        \centering
        \includegraphics[width=\linewidth]{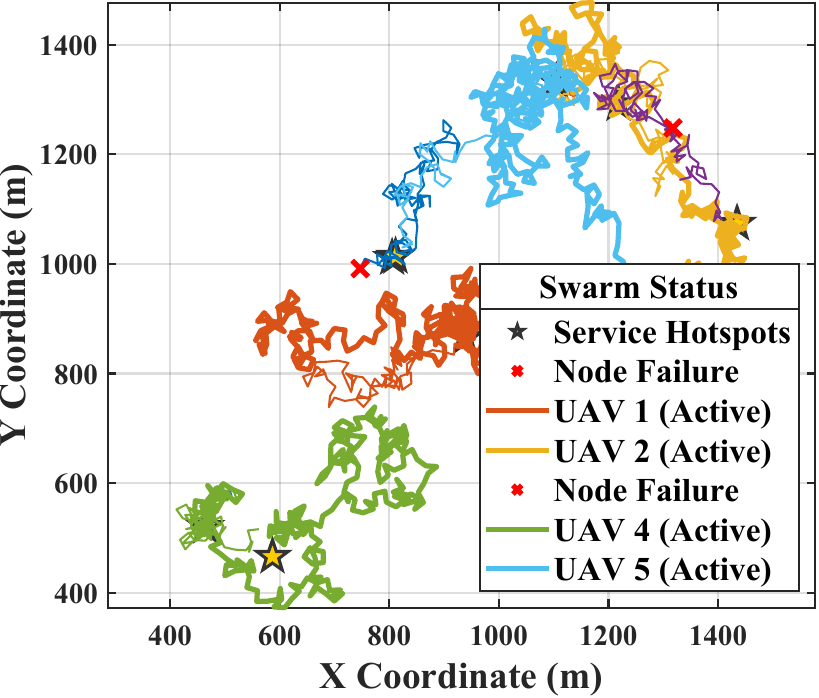}
        \caption{Swarm Trajectory Reconfiguration}
        \label{fig:mechanism_resilience:traj}
    \end{subfigure}\hfill
    \begin{subfigure}[b]{0.49\columnwidth}
        \centering
        \includegraphics[width=\linewidth]{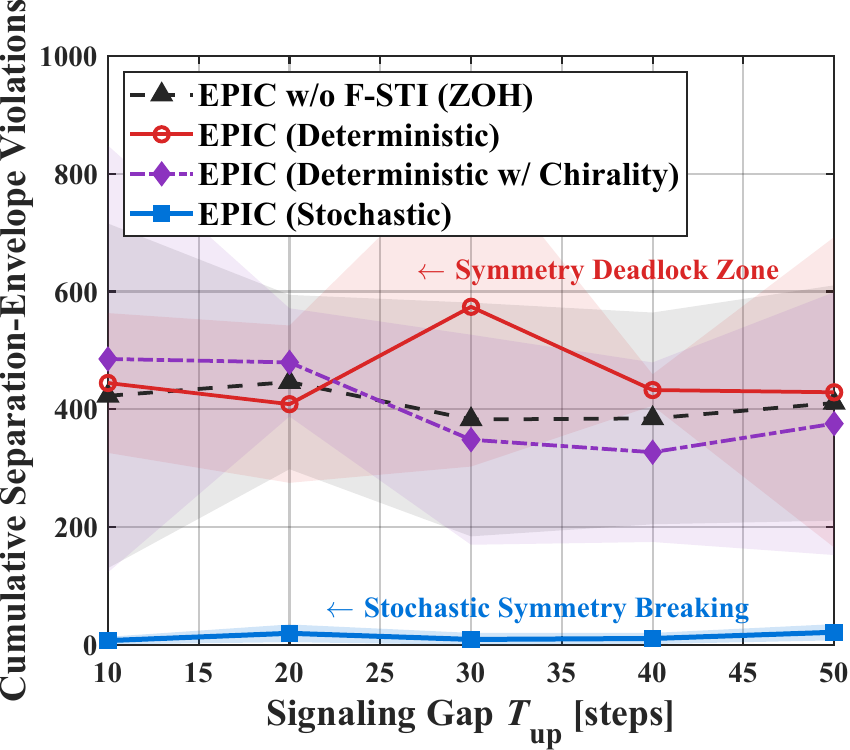}
        \caption{Safety Sensitivity vs. Signaling Gap}
        \label{fig:mechanism_resilience:collision}
    \end{subfigure}
    \caption{Swarm coordination robustness and safety sensitivity under catastrophic operational conditions. (a) Autonomous spatial reconfiguration filling coverage vacancies after $u_0$ and $u_3$ fail at $t = 200$ ($T_{\mathrm{up}} = 30$). (b) Cumulative separation-envelope violations ($<30$~m separation envelope) versus signaling gap $T_{\mathrm{up}} \in [10, 50]$ steps, demonstrating that stochastic symmetry breaking substantially outperforms deterministic execution, kinematic-free ZOH, and rule-based geometric chirality. Shaded bands denote standard deviations across independent evaluation seeds.}
    \label{fig:mechanism_resilience}
\end{figure}

Building upon this cross-task generalization, we stress-test EPIC under catastrophic node failures and varying communication latencies to isolate the mechanisms governing swarm resilience. In Fig.~\ref{fig:mechanism_resilience:traj}, two agents ($u_0, u_3$) are abruptly deactivated at $t = 200$ under an extended signaling gap of $T_{\mathrm{up}} = 30$ steps. Without external coordination, the surviving agents autonomously reconfigure their spatial coverage to absorb the orphaned service hotspots prior to central reconciliation; the observed trajectory variations are consistent with the symmetry-breaking effect induced by stochastic action sampling.

Fig.~\ref{fig:mechanism_resilience:collision} systematically evaluates this safety resilience across signaling gaps $T_{\mathrm{up}} \in [10, 50]$ steps against three comparative baselines. Without action stochasticity, EPIC (Deterministic) collapses into an extensive \textit{Symmetry Deadlock Zone} ($400 \sim 580$ breaches) as symmetric agents repeatedly pursue the same deficit centers. Augmenting deterministic actions with an ad-hoc geometric filter (EPIC Deterministic w/ Chirality, enforcing a fixed right-hand avoidance bias) yields negligible improvement ($330 \sim 480$ breaches), demonstrating that weak local repulsions are readily overpowered by dominant macroscopic target gradients during congested encounters. Similarly, the kinematic-free baseline, EPIC w/o F-STI (ZOH), incurs heavy open-loop extrapolation drift ($380 \sim 450$ breaches). In contrast, EPIC (Stochastic) maintains robust trajectory diversification, suppressing envelope violations below 25 across all outage horizons ($<0.28\%$ event rate, a $>90\%$ reduction in separation-envelope violations relative to the deterministic regimes). These results indicate that stochastic execution is effective for mitigating macro-level deadlocks under extended signaling voids, while residual separation-envelope violations may require a dedicated high-frequency physical safety layer in safety-critical deployments.

\subsubsection{Micro-Analysis of Perception-Action Robustness}
\label{ssbsec:micro_analysis}

\begin{figure}[!t]
    \centering
    \begin{subfigure}[b]{0.49\columnwidth}
        \centering
        \includegraphics[width=\linewidth]{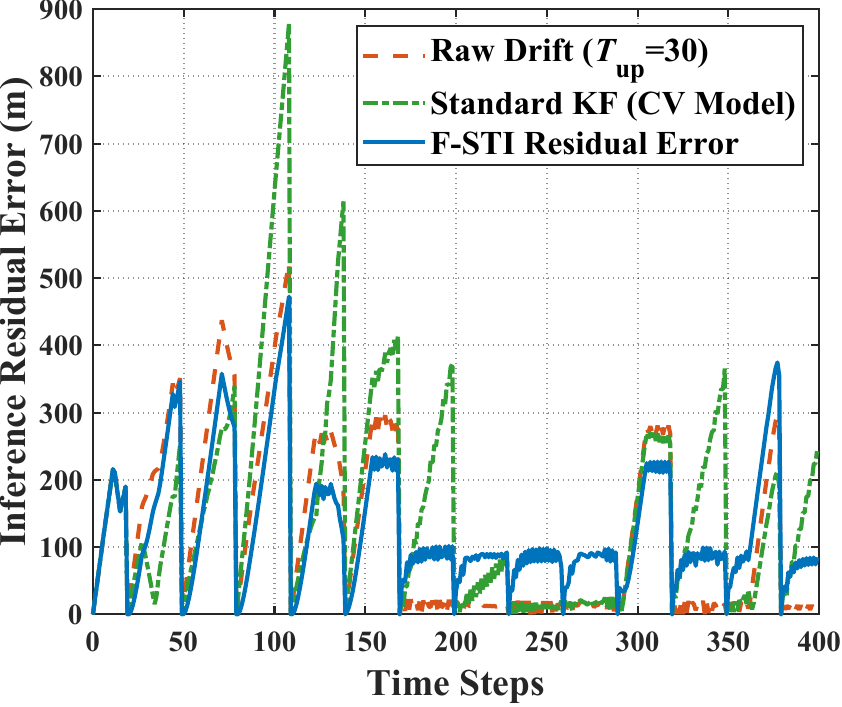}
        \caption{Residual Error}
        \label{fig:inference:fsti_analysis}
    \end{subfigure}\hfill
    \begin{subfigure}[b]{0.49\columnwidth}
        \centering
        \includegraphics[width=\linewidth]{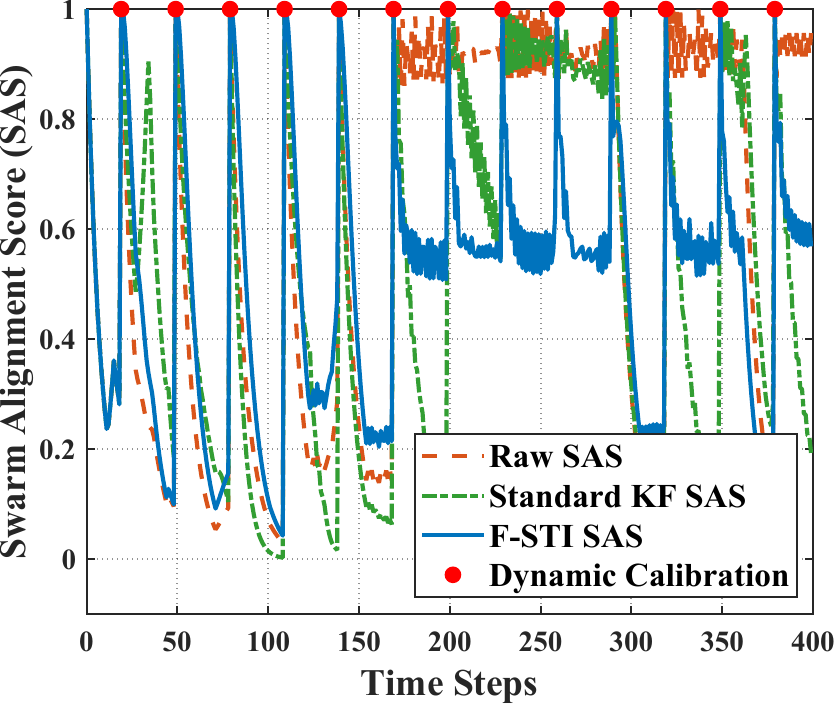}
        \caption{SAS Alignment}
        \label{fig:inference:sas_analysis}
    \end{subfigure}    
    \caption{Micro-analysis under extreme asynchronous signaling ($T_{\mathrm{up}}=30$). (a) Residual tracking errors of raw drift, standard CV-KF, and F-STI. (b) Swarm Alignment Score (SAS) under the same estimators, with red markers indicating successful peer-state synchronization.}
    \label{fig:inference}
\end{figure}

To evaluate state-reconstruction robustness under prolonged signaling gaps, Fig.~\ref{fig:inference} compares conventional KF-based prediction with the proposed F-STI under the $T_{\mathrm{up}}=30$ stress test.

First, Fig.~\ref{fig:inference:fsti_analysis} compares the residual tracking errors of a standard KF with a \textit{constant-velocity} (CV) model \cite{BarShalom2002} and the proposed F-STI. Under $\delta t=1.0$~s and $v_{\max}=15.0$~m/s, the uncompensated open-loop displacement can reach $30\times15.0=450$~m over a 30-step silence interval. The CV-KF exhibits larger and more variable error excursions, whereas F-STI maintains substantially lower residual errors across most outage intervals.

The corresponding SAS results in Fig.~\ref{fig:inference:sas_analysis} show the same trend in trajectory fidelity. Periodic recovery results from successful peer-state synchronization, which resets the prediction error. During signaling silences, F-STI maintains higher alignment than both the raw and KF-based predictions.

\subsubsection{Signaling-Gap Robustness and Response Analysis}
\label{ssbsec:latency_staleness}

We examine how signaling gaps affect coordination timeliness and task performance. Here, \textit{response delay} (or \textit{information-to-action response delay}) characterizes the elapsed time between an unobserved state transition and the swarm's tactical adaptation under intermittent connectivity, evaluated per decision epoch rather than by the discrete control-step duration $\delta t=1.0$~s.

\begin{figure}[!t]
    \centering
    \begin{subfigure}[b]{0.49\columnwidth}
        \centering
        \includegraphics[width=\linewidth]{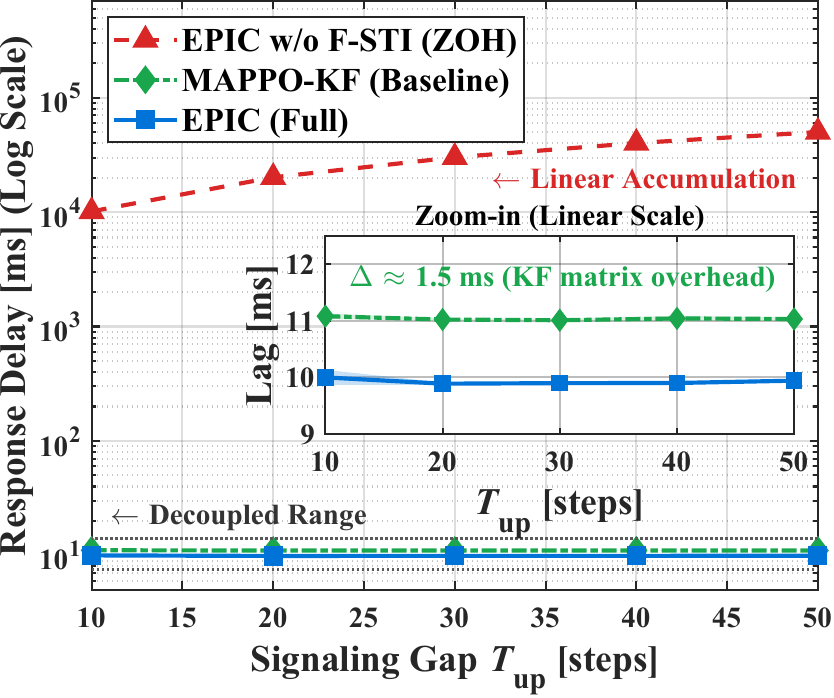}
        \caption{Response Delay vs. Signaling Gap}
        \label{fig:latency_staleness:response_delay}
    \end{subfigure}\hfill
    \begin{subfigure}[b]{0.49\columnwidth}
        \centering
        \includegraphics[width=\linewidth]{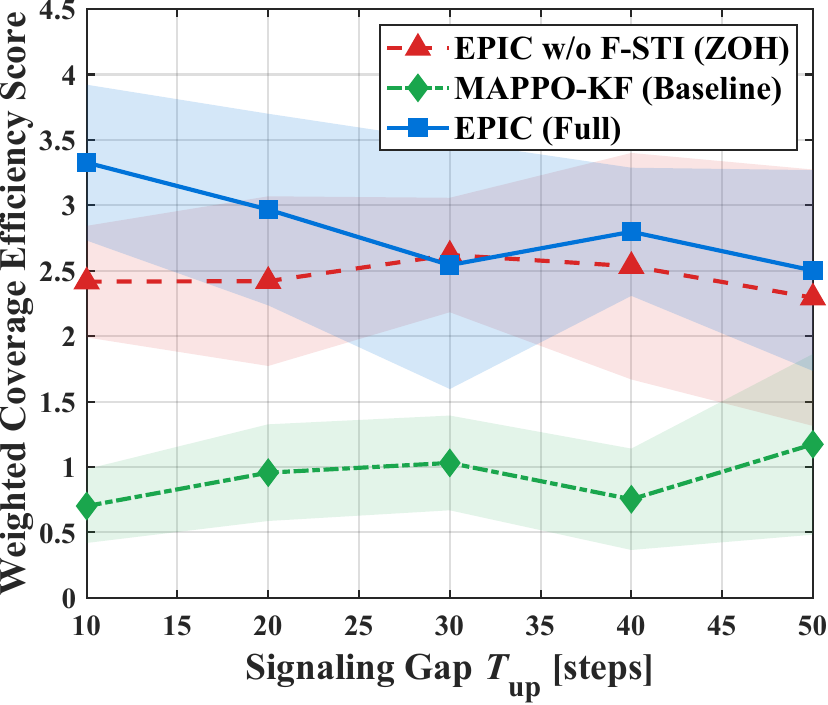}
        \caption{WCE Score vs. Signaling Gap}
        \label{fig:latency_staleness:wce_staleness}
    \end{subfigure}
    \caption{Impact of signaling gaps on response delay and task robustness: (a) information-to-action response delay; (b) steady-state WCE. Shaded regions denote standard deviations across independent evaluation environments.}
    \label{fig:latency_staleness}
\end{figure}

As shown in Fig.~\ref{fig:latency_staleness:response_delay}, EPIC w/o F-STI (ZOH) exhibits an approximately linear increase in response delay, exceeding $5\times10^{4}$~ms at $T_{\mathrm{up}}=50$, as unavailable peer states remain frozen until the next physical update. In contrast, EPIC (Full) and MAPPO-KF maintain nearly constant information-to-action response delays of $\sim10$~ms across $T_{\mathrm{up}}\in[10,50]$, showing that local state reconstruction enables action generation without waiting for the next peer-state update. The inset highlights an approximately $1.5$~ms matrix-inversion and covariance overhead for MAPPO-KF, while the measured per-cycle decision latency is approximately $11.0$~ms for MAPPO-KF versus $9.9$~ms for EPIC with closed-form F-STI. All per-cycle latency measurements were obtained on a PC equipped with an AMD Ryzen 9 9950X CPU under a single-process, single-core execution setting.

\begin{remark}[Response-Time Robustness and Computational Overhead]
    \label{rmk:realtime_feasibility}
    EPIC and MAPPO-KF maintain nearly constant response delays as the signaling gap increases, whereas the reactive ZOH baseline experiences progressively larger delays. Moreover, F-STI avoids the explicit matrix-inversion operation required by KF, reducing the computational overhead of local state reconstruction while preserving predictive continuity.
\end{remark}

Low response delay alone, however, does not guarantee effective coordination. As shown in Fig.~\ref{fig:latency_staleness:wce_staleness}, MAPPO-KF maintains low response delay but achieves lower steady-state WCE scores ($0.7\sim1.2$), while EPIC w/o F-STI (ZOH) achieves $2.3\sim2.6$ and EPIC (Full) attains $2.5\sim3.3$. Together with Fig.~\ref{fig:inference}, these results show that resilient coordination requires both timely response and accurate reconstruction of unavailable peer states.

\subsubsection{Resilience Performance and Ablation Analysis}
\label{ssbsec:resilience_ablation}

\begin{figure}[!t]    
    \centering
    \begin{subfigure}[b]{0.48\columnwidth}
        \centering
        \includegraphics[width=\linewidth]{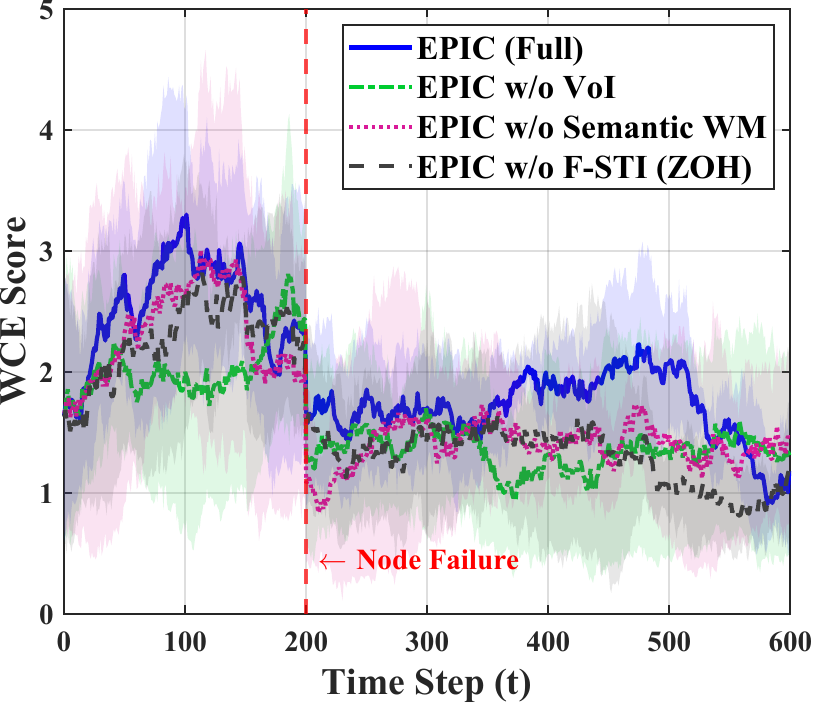}
        \caption{WCE Score Recovery}
        \label{fig:resilience:wce}
    \end{subfigure}\hfill
    \begin{subfigure}[b]{0.495\columnwidth}
        \centering
        \includegraphics[width=\linewidth]{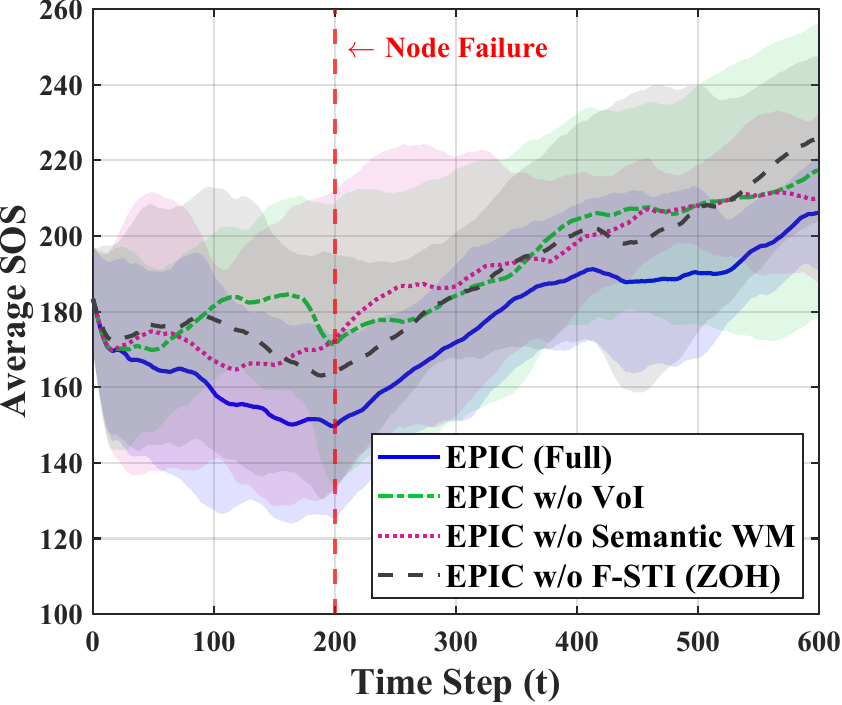}
        \caption{SOS Recovery Performance}
        \label{fig:resilience:sos}
    \end{subfigure}     
    \caption{Resilience and ablation analysis across five independent seeds. (a) WCE recovery profiles comparing EPIC (Full), ablation variants, and EPIC w/o F-STI (ZOH). (b) SOS performance illustrating the benefit of VoI-triggered synchronization.}
    \label{fig:resilience}
\end{figure}

Building upon the stable performance under high-latency conditions, we evaluate quantitative recovery under catastrophic node failures in Fig.~\ref{fig:resilience} across five independent seeds. As shown in Fig.~\ref{fig:resilience:wce}, EPIC w/o F-STI (ZOH) lacks predictive awareness and exhibits a pronounced reduction in steady-state WCE following the node failure at $t=200$.

EPIC w/o VoI and EPIC w/o Semantic WM retain the F-STI operator, while the latter removes the semantic world-model component; therefore, both variants maintain physical tracking initially, but their long-term performance degrades and approaches the EPIC w/o F-STI (ZOH) baseline in the later stage (e.g., after $t=400$). This indicates that proactive state extrapolation alone is insufficient for sustained recovery: removing VoI-triggered synchronization or the semantic world model respectively limits adaptive realignment and task-aware coordination.

In contrast, EPIC (Full) achieves the most consistent WCE recovery in Fig.~\ref{fig:resilience:wce} while maintaining the lowest and most stable SOS in Fig.~\ref{fig:resilience:sos}. The results demonstrate that combining F-STI, semantic world modeling, and VoI-triggered synchronization provides complementary benefits for resilient swarm coordination after catastrophic disruptions.

\subsection{Multi-Time Scale Sensitivity and Stability Analysis}
\label{subsec:sensitivity_scalability}

We further examine the sensitivity of EPIC to the synchronization time scales that govern local calibration and global federated aggregation. Specifically, $T_{\mathrm{cal}}$ represents the interval between local semantic recalibrations, whereas $T_{\mathrm{agg}}$ denotes the scheduled global aggregation interval. By extending the evaluation horizon to 2,400 steps, this study evaluates whether the adopted synchronization settings require precise tuning to maintain stable task performance.

\begin{figure}[!t]
    \centering
    \begin{subfigure}[b]{0.24\textwidth}
        \centering
        \includegraphics[width=\linewidth]{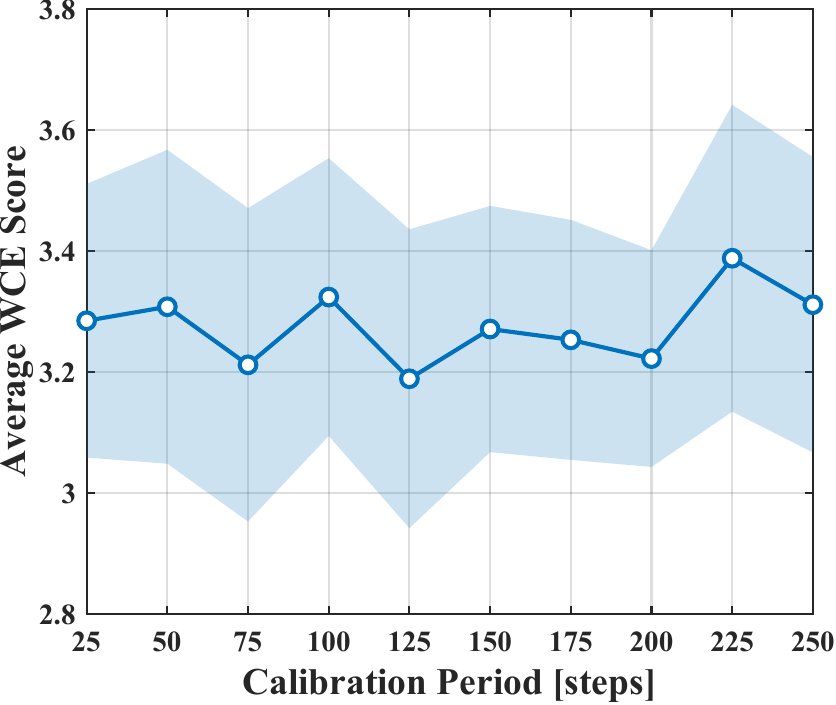}
        \caption{Sensitivity to $T_{\mathrm{cal}}$}
        \label{fig:sensitivity:cal}
    \end{subfigure}
    \hfill
    \begin{subfigure}[b]{0.24\textwidth}
        \centering
        \includegraphics[width=\linewidth]{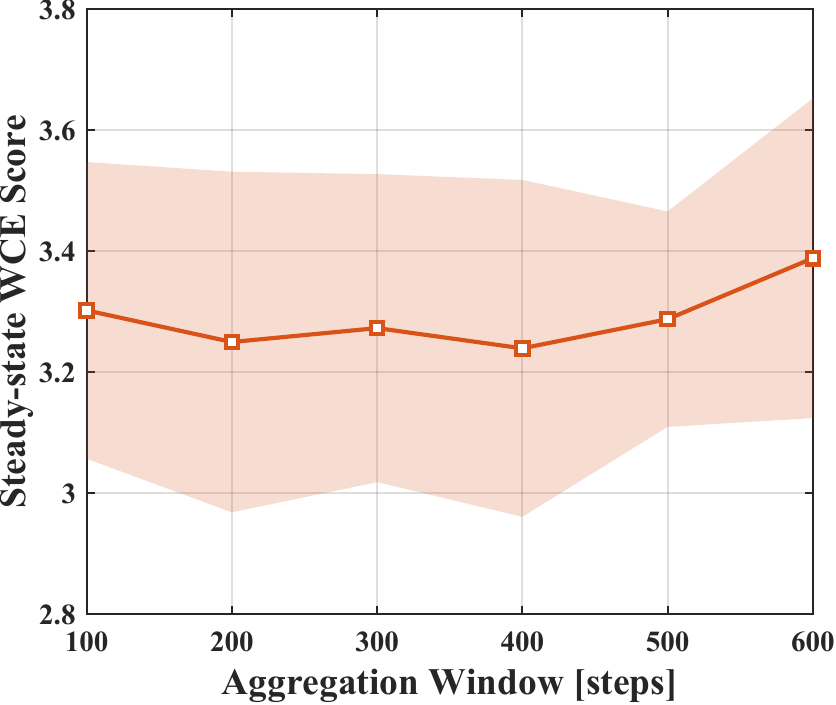}
        \caption{Sensitivity to $T_{\mathrm{agg}}$}
        \label{fig:sensitivity:agg}
    \end{subfigure}\\ 
    \vspace{.5em}
    \begin{subfigure}[b]{0.24\textwidth}
        \centering
        \includegraphics[width=\linewidth]{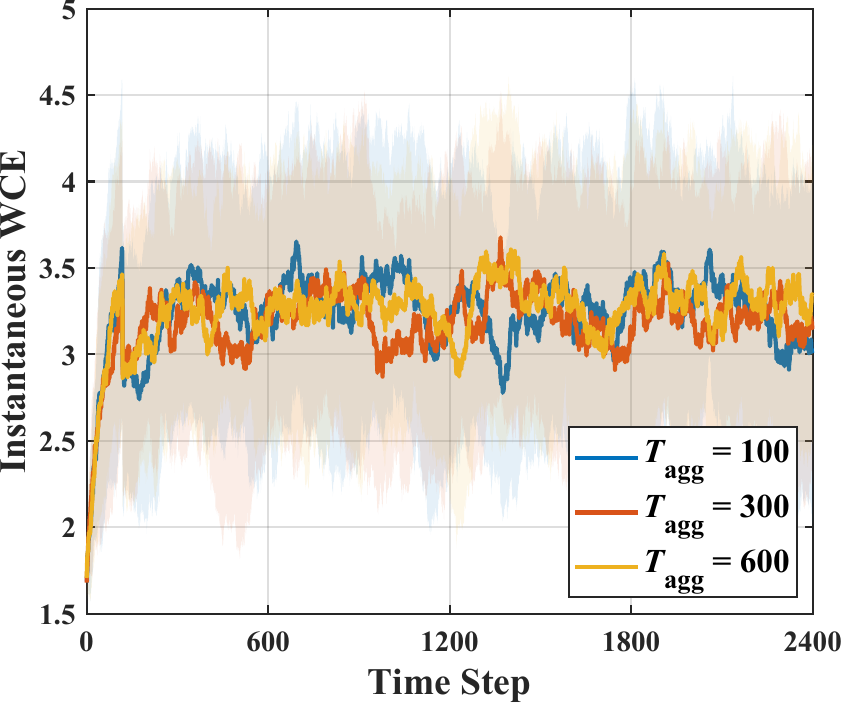}
        \caption{Temporal WCE trajectories}
        \label{fig:sensitivity:agg_convergence}
    \end{subfigure}        
    \caption{Multi-time scale sensitivity analysis over 2,400 steps.
    (a) Steady-state WCE across local calibration periods $T_{\mathrm{cal}}$.
    (b) Steady-state WCE across global aggregation windows $T_{\mathrm{agg}}$.
    (c) Temporal evolution of instantaneous WCE for representative
    $T_{\mathrm{agg}}$ configurations, illustrating empirical robustness
    to synchronization granularity.}
    \label{fig:sensitivity}
\end{figure}

As illustrated in Fig.~\ref{fig:sensitivity:cal}, the steady-state WCE varies only modestly as $T_{\mathrm{cal}}$ increases from 25 to 250 steps. This indicates that EPIC retains effective local state and semantic consistency over a relatively broad range of calibration intervals, rather than relying on a narrowly tuned value. Similarly, Fig.~\ref{fig:sensitivity:agg} shows only a limited change in the steady-state WCE as $T_{\mathrm{agg}}$ varies from 100 to 600 steps. A larger aggregation interval introduces a modest performance degradation, consistent with the reduced frequency of global model refreshes, but the overall performance remains within a relatively stable regime across the evaluated range.

To examine the long-horizon behavior under sparse global updates, Fig.~\ref{fig:sensitivity:agg_convergence} reports the instantaneous WCE over 2,400 steps for representative $T_{\mathrm{agg}}$ configurations. While smaller aggregation intervals provide slightly faster initial adaptation during the first 600 steps, the evaluated configurations subsequently approach similar steady-state performance regimes. No pronounced divergence or persistent oscillation is observed between aggregation events, supporting the empirical robustness of EPIC to the tested synchronization granularities.

Overall, these results suggest that the performance of EPIC is relatively insensitive to the synchronization time scales within the evaluated ranges. Accordingly, we adopt $T_{\mathrm{cal}}=100$ and $T_{\mathrm{agg}}=300$ steps as representative mid-range operating points, rather than claiming them to be globally optimal settings. This sensitivity study is intended to characterize the practical robustness of the adopted synchronization design; it does not constitute a separate convergence analysis of federated learning under arbitrary asynchronous scheduling and task distributions.

\subsection{Scalability Stress Tests}
\label{subsec:scalability_analysis}

To ensure a fair comparison across different swarm scales, each configuration is evaluated over 1,200 steps. We report the steady-state WCE score by averaging the performance from $t=201$ to $t=1,200$, thereby reducing the impact of the initial transient navigation phase and focusing on the long-term coordination capability of the swarm.

To evaluate the operational robustness of EPIC in increasingly complex environments, we conduct scalability tests by varying the number of UAVs $N$ and the number of hotspots $K$. These two dimensions respectively characterize increasing swarm density and mission complexity, allowing us to examine whether the learned policy maintains effective coordination as the operational scale expands.

\begin{figure}[!t]
    \centering
    \begin{subfigure}[b]{0.49\columnwidth}
        \centering
        \includegraphics[width=\linewidth]{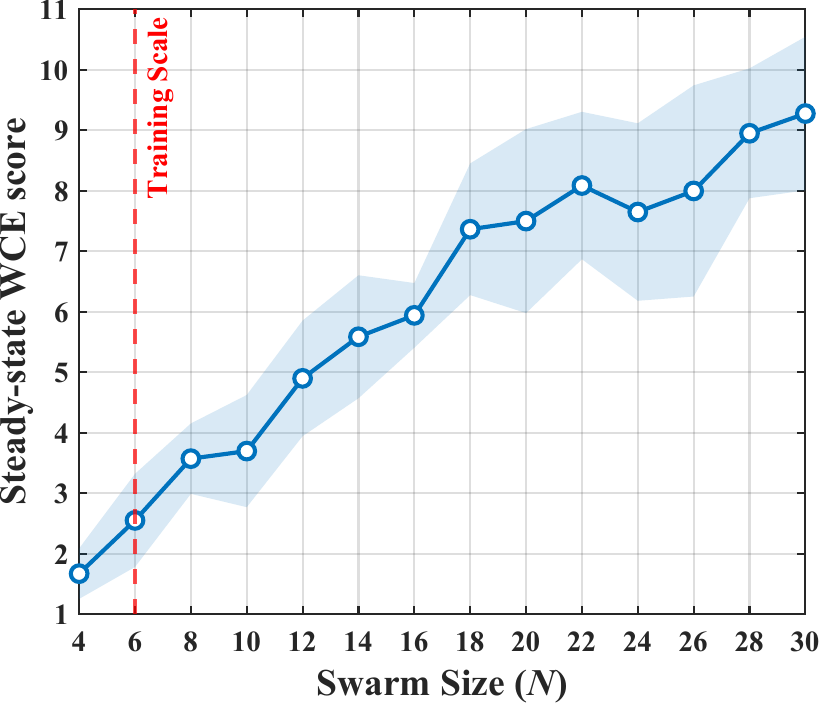}
        \caption{Swarm size scalability}
        \label{fig:scalability:swarm_size}
    \end{subfigure}
    \hfill
    \begin{subfigure}[b]{0.49\columnwidth}
        \centering
        \includegraphics[width=\linewidth]{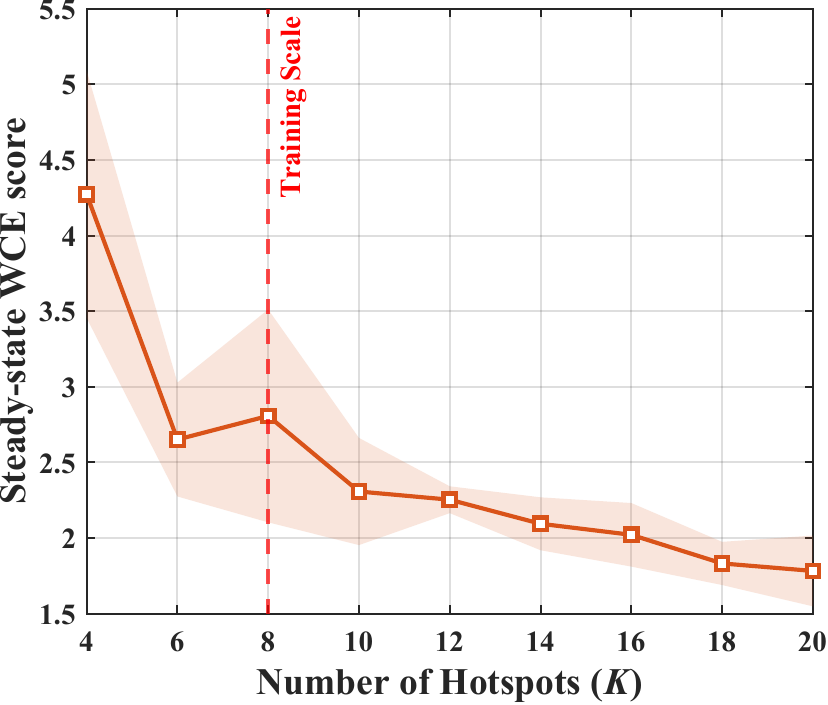}
        \caption{Hotspot density scalability}
        \label{fig:scalability:hotspot_density}
    \end{subfigure}    
    \caption{Scalability analysis. (a) Steady-state WCE versus swarm size $N$ ($K=8$). (b) Steady-state WCE versus hotspot density $K$ ($N=6$). Dashed red lines denote training configurations.}
    \label{fig:scalability}
\end{figure}

As illustrated in Fig.~\ref{fig:scalability:swarm_size}, the swarm size $N$ is increased from 4 to 30 while keeping the mission area and $K=8$ fixed. The steady-state WCE increases from approximately 1.8 to 9.2 as additional UAVs are introduced, without evident oscillatory behavior or abrupt performance degradation. This indicates that the added agents continue to contribute to mission coverage over the evaluated range. Notably, the learned coordination policy remains effective beyond the training configuration ($N=6$), indicating stable task performance over the evaluated swarm-size range.

Conversely, Fig.~\ref{fig:scalability:hotspot_density} examines increasing task density by varying $K$ from 4 to 20 while fixing $N=6$. The steady-state WCE decreases monotonically as the fixed number of UAVs must distribute their coverage capability across a larger number of competing hotspots, reflecting the expected resource-dilution effect. Nevertheless, the decline remains smooth, with the WCE staying around 1.8 even at $K=20$, more than twice the training configuration. This result indicates that EPIC maintains a stable coordination baseline as mission complexity increases over the evaluated range.

\section{Conclusion}
\label{sec:conclusion}
This paper proposed EPIC, a communication-aware digital twin framework for resilient UAV swarm coordination in wireless-enabled LAE networks. By integrating a VoI-triggered F-STI operator with Federated MARL, EPIC enables agents to maintain an Embodied Semantic World Model for proactive inference during signaling silences, thereby mitigating open-loop kinematic drift and preserving decision robustness. The theoretical analysis characterizes the decentralized interaction as a state-conditional congestion game, whose potential structure provides an equilibrium reference under canonical action discretization. Under intermittent connectivity, we further derive an $\epsilon$-Nash robustness bound for an equilibrium of the perceived game when state uncertainty remains bounded under the stated regularity conditions. Extensive evaluations demonstrate a nominal 66.7\% reduction in scheduled uplink frequency while sustaining robust service coverage under 30-step communication voids. Overall, EPIC provides an effective balance between communication efficiency and coordination robustness for intermittent aerial networks. Future work will further assess EPIC through hardware-in-the-loop experiments and real-flight validation.

\bibliographystyle{IEEEtran}
\bibliography{IEEEabrv,reference}

\ifCLASSOPTIONcaptionsoff  \newpage \fi

\end{document}

\maketitle

\section{Purpose and Scope}

This supplement provides the complete mathematical derivations underlying the theoretical analysis of the EPIC framework in the revised manuscript. The main manuscript presents the essential statements and compact proofs; here, the intermediate steps are expanded to make the assumptions, inequalities, and logical dependencies explicit.

The analysis follows the same logical organization as the revised manuscript.
The arrows below indicate the analytical progression rather than a direct
mathematical implication between every consecutive component:

\begin{center}
\begin{tabular}{c}
\textbf{Kinematic Error and Semantic Robustness}
\\[-0.05em]
{\small Finite-Horizon Kinematic Inference Bound and Semantic Misclassification Condition}
\\[0.35em]
$\Downarrow$
\\[0.35em]
\textbf{Bounded Semantic-Model Disagreement Relative to Federated Anchor}
\\[-0.05em]
{\small Federated Semantic-Model Alignment}
\\[0.35em]
$\Downarrow$
\\[0.35em]
\textbf{Cooperative Coverage Objective and State-Conditional Congestion Game}
\\[-0.05em]
{\small Exact-Potential Equilibrium Reference}
\\[0.35em]
$\Downarrow$
\\[0.35em]
\textbf{Approximate Equilibrium under Intermittent Connectivity}
\\[-0.05em]
{\small Composite State Uncertainty and Uniform Stage-Utility Perturbation}
\\[0.35em]
$\Downarrow$
\\[0.35em]
\textbf{$\epsilon$-Nash Robustness}
\end{tabular}
\end{center}

The first analytical block connects finite-horizon F-STI inference error to semantic coverage robustness. The second bounds anchor-relative semantic-model disagreement under federated calibration, while the third establishes the complete-information exact-potential equilibrium reference. The final block combines these ingredients through a bounded state-uncertainty and utility-perturbation analysis to establish the resulting $\epsilon$-Nash robustness.

The kinematic and perturbation results are deliberately finite-horizon, while the game-theoretic statements are state-conditional. The calibration recurrence additionally admits the conditioned steady-state bound in Proposition~\ref{prop:anchor_disagreement} under the fixed-anchor analysis; this should not be interpreted as a general convergence result for the operational asynchronous federated-learning process. In particular, the potential-game result is not a convergence claim for MAPPO training, and the $\epsilon$-Nash result is a perturbation transfer result for a perceived stage-game equilibrium.

\section{Kinematic Error and Semantic Robustness}

\subsection{Discrete-Time F-STI Dynamics}

Consider a signaling silence beginning at time $t$. Let
$\delta t>0$ denote the physical control interval and let
\[
k\in\{0,1,\ldots,T_{\mathrm{up}}\}
\]
be the number of elapsed discrete steps. The true kinematics are
\begin{equation}
\mathbf{p}_{t+\ell+1}
=
\mathbf{p}_{t+\ell}
+
\mathbf{v}_{t+\ell}\delta t,
\end{equation}
and
\begin{equation}
\mathbf{v}_{t+\ell+1}
=
\mathbf{v}_{t+\ell}
+
\mathbf{a}_{t+\ell}\delta t.
\end{equation}
The F-STI velocity predictor is
\begin{equation}
\hat{\mathbf{v}}_{t+\ell}
=
(1-\alpha)^\ell \hat{\mathbf{v}}_t,
\qquad
0\le\alpha\le1,
\end{equation}
with inferred position propagated using
\begin{equation}
\hat{\mathbf{p}}_{t+\ell+1}
=
\hat{\mathbf{p}}_{t+\ell}
+
\hat{\mathbf{v}}_{t+\ell}\delta t.
\end{equation}
At a successful synchronization instant,
\begin{equation}
\mathbf{p}_t=\hat{\mathbf{p}}_t.
\end{equation}

\begin{assumption}[Bounded true acceleration]
During the signaling silence interval,
\begin{equation}
\|\mathbf{a}_{t+\ell}\|\le a_{\max}
\end{equation}
for every admissible $\ell$.
\end{assumption}

\begin{lemma}[Finite-Horizon Kinematic Inference Bound]\label{lemma:drift}
For every
$k\in\{0,\ldots,T_{\mathrm{up}}\}$,
the spatial inference error
\[
\epsilon_k
=
\|\mathbf{p}_{t+k}-\hat{\mathbf{p}}_{t+k}\|
\]
satisfies
\begin{equation}
\label{eq:supp_drift_bound}
\epsilon_k
\le
\delta t
\sum_{\ell=0}^{k-1}
\left[
\|\mathbf{v}_t-\hat{\mathbf{v}}_t\|
+
a_{\max}\ell\delta t
+
\|\hat{\mathbf{v}}_t\|
\left(1-(1-\alpha)^\ell\right)
\right].
\end{equation}
\end{lemma}

\begin{proof}
We first derive a bound on the velocity mismatch.

From the true velocity recursion,
\begin{equation}
\mathbf{v}_{t+\ell}
=
\mathbf{v}_t
+
\sum_{r=0}^{\ell-1}
\mathbf{a}_{t+r}\delta t.
\end{equation}
Hence
\begin{equation}
\mathbf{v}_{t+\ell}-\hat{\mathbf{v}}_{t+\ell}
=
\left(\mathbf{v}_t-\hat{\mathbf{v}}_t\right)
+
\sum_{r=0}^{\ell-1}\mathbf{a}_{t+r}\delta t
+
\hat{\mathbf{v}}_t-\hat{\mathbf{v}}_{t+\ell}.
\end{equation}
Because
\[
\hat{\mathbf{v}}_{t+\ell}
=
(1-\alpha)^\ell\hat{\mathbf{v}}_t,
\]
the last term is
\begin{equation}
\hat{\mathbf{v}}_t-\hat{\mathbf{v}}_{t+\ell}
=
\left[1-(1-\alpha)^\ell\right]\hat{\mathbf{v}}_t.
\end{equation}
Taking norms and applying the triangle inequality gives
\begin{align}
\|\mathbf{v}_{t+\ell}-\hat{\mathbf{v}}_{t+\ell}\|
&\le
\|\mathbf{v}_t-\hat{\mathbf{v}}_t\|
+
\left\|
\sum_{r=0}^{\ell-1}\mathbf{a}_{t+r}\delta t
\right\|
+
\|\hat{\mathbf{v}}_t\|
\left[1-(1-\alpha)^\ell\right]
\nonumber\\
&\le
\|\mathbf{v}_t-\hat{\mathbf{v}}_t\|
+
\sum_{r=0}^{\ell-1}
\|\mathbf{a}_{t+r}\|\delta t
+
\|\hat{\mathbf{v}}_t\|
\left[1-(1-\alpha)^\ell\right]
\nonumber\\
&\le
\|\mathbf{v}_t-\hat{\mathbf{v}}_t\|
+
a_{\max}\ell\delta t
+
\|\hat{\mathbf{v}}_t\|
\left[1-(1-\alpha)^\ell\right].
\label{eq:supp_velocity_bound}
\end{align}
The condition $0\le\alpha\le1$ guarantees
$0\le(1-\alpha)^\ell\le1$, so the final damping term is nonnegative.

We next propagate the velocity error into position error. Define
\[
\mathbf{e}_{p,\ell}
=
\mathbf{p}_{t+\ell}-\hat{\mathbf{p}}_{t+\ell}.
\]
Using the discrete position recursions,
\begin{equation}
\mathbf{e}_{p,\ell+1}
=
\mathbf{e}_{p,\ell}
+
\left(
\mathbf{v}_{t+\ell}-\hat{\mathbf{v}}_{t+\ell}
\right)\delta t.
\end{equation}
Unrolling this recursion gives
\begin{equation}
\mathbf{e}_{p,k}
=
\mathbf{e}_{p,0}
+
\delta t
\sum_{\ell=0}^{k-1}
\left(
\mathbf{v}_{t+\ell}-\hat{\mathbf{v}}_{t+\ell}
\right).
\end{equation}
Since the position is verified at the synchronization instant,
$\mathbf{e}_{p,0}=\mathbf{0}$. Thus,
\begin{align}
\epsilon_k
&=
\|\mathbf{e}_{p,k}\|
\nonumber\\
&\le
\delta t
\sum_{\ell=0}^{k-1}
\|\mathbf{v}_{t+\ell}-\hat{\mathbf{v}}_{t+\ell}\|.
\end{align}
Substituting~\eqref{eq:supp_velocity_bound} yields~\eqref{eq:supp_drift_bound}.
\end{proof}

\subsection{Effect of Boundary Clipping}

The implementation uses the projection
\[
\hat{\mathbf p}_{t+\ell+1}
=
\Pi_{\Omega}
\left(
\hat{\mathbf p}_{t+\ell}
+
\hat{\mathbf v}_{t+\ell}\delta t
\right),
\]
where the operational region $\Omega=[0,L]^2$ and
$\Pi_{\Omega}$ is Euclidean projection onto $\Omega$.

Because Euclidean projection onto a closed convex set is nonexpansive,
\begin{equation}
\|\Pi_{\Omega}(\mathbf x)-\Pi_{\Omega}(\mathbf y)\|
\le
\|\mathbf x-\mathbf y\|.
\end{equation}
For an admissible true position
$\mathbf p_{t+\ell+1}\in\Omega$,
$\Pi_{\Omega}(\mathbf p_{t+\ell+1})=\mathbf p_{t+\ell+1}$.
Therefore, even when the inferred trajectory encounters an operational
boundary, projection cannot increase the discrepancy:
\begin{align}
\|\mathbf p_{t+\ell+1}-\hat{\mathbf p}_{t+\ell+1}\|
&=
\left\|
\Pi_{\Omega}(\mathbf p_{t+\ell+1})
-
\Pi_{\Omega}
\left(
\hat{\mathbf p}_{t+\ell}+
\hat{\mathbf v}_{t+\ell}\delta t
\right)
\right\|
\nonumber\\
&\le
\left\|
\mathbf p_{t+\ell+1}
-
\hat{\mathbf p}_{t+\ell}
-
\hat{\mathbf v}_{t+\ell}\delta t
\right\|.
\end{align}
Hence the clipping operation does not enlarge the kinematic error bound.
The main derivation can therefore be interpreted as a conservative
upper bound for both the unclipped recursion and its projected
implementation.

\subsection{Semantic Coverage Misclassification}

\begin{proposition}[Semantic Misclassification Condition]
\label{prop:semantic_error}
Let
\begin{equation}
\chi_{i,m}(\mathbf p)
=
\mathbb{1}
\left\{
\|\mathbf p-\mathbf g_m\|
\le R_{\mathrm{cov}}
\right\}
\end{equation}
denote the binary coverage status of grid $m$. If
\[
\|\mathbf p-\hat{\mathbf p}\|\le\epsilon_k,
\]
then a classification mismatch
\[
\chi_{i,m}(\mathbf p)
\ne
\chi_{i,m}(\hat{\mathbf p})
\]
can occur only when
\begin{equation}
\label{eq:supp_semantic_condition}
\left|
\|\hat{\mathbf p}-\mathbf g_m\|
-
R_{\mathrm{cov}}
\right|
\le
\epsilon_k.
\end{equation}
\end{proposition}

\begin{proof}
By the reverse triangle inequality,
\begin{align}
\left|
\|\mathbf p-\mathbf g_m\|
-
\|\hat{\mathbf p}-\mathbf g_m\|
\right|
&\le
\|\mathbf p-\hat{\mathbf p}\|
\nonumber\\
&\le
\epsilon_k.
\label{eq:supp_reverse_triangle}
\end{align}

Suppose first that
\[
\|\hat{\mathbf p}-\mathbf g_m\|
>
R_{\mathrm{cov}}+\epsilon_k.
\]
Then~\eqref{eq:supp_reverse_triangle} implies
\begin{align}
\|\mathbf p-\mathbf g_m\|
&\ge
\|\hat{\mathbf p}-\mathbf g_m\|-\epsilon_k
>
R_{\mathrm{cov}},
\end{align}
so both true and inferred positions are outside the coverage region:
\[
\chi_{i,m}(\mathbf p)
=
\chi_{i,m}(\hat{\mathbf p})
=
0.
\]

Conversely, suppose that
\[
\|\hat{\mathbf p}-\mathbf g_m\|
<
R_{\mathrm{cov}}-\epsilon_k.
\]
Then
\begin{align}
\|\mathbf p-\mathbf g_m\|
&\le
\|\hat{\mathbf p}-\mathbf g_m\|+\epsilon_k
<
R_{\mathrm{cov}},
\end{align}
and therefore
\[
\chi_{i,m}(\mathbf p)
=
\chi_{i,m}(\hat{\mathbf p})
=
1.
\]

Thus, a mismatch can only occur when neither deterministic condition
above holds, namely when
\[
R_{\mathrm{cov}}-\epsilon_k
\le
\|\hat{\mathbf p}-\mathbf g_m\|
\le
R_{\mathrm{cov}}+\epsilon_k,
\]
which is equivalent to~\eqref{eq:supp_semantic_condition}.
\end{proof}

The interval
\[
[R_{\mathrm{cov}}-\epsilon_k,\,
 R_{\mathrm{cov}}+\epsilon_k]
\]
is a conservative enclosing band of width $2\epsilon_k$ around the
coverage boundary. Since a physical radial distance is nonnegative, the
feasible ambiguity region is its intersection with $[0,\infty)$, i.e.,
\[
\left[
\max\{0,R_{\mathrm{cov}}-\epsilon_k\},\,
R_{\mathrm{cov}}+\epsilon_k
\right].
\]
Thus, when $\epsilon_k>R_{\mathrm{cov}}$, the feasible ambiguity region
is no larger than the $2\epsilon_k$ conservative enclosing band. This
clarification preserves the sufficient misclassification condition in
\eqref{eq:supp_semantic_condition} without assuming that the entire
formal radial interval is physically realizable.

\section{Bounded Semantic-Model Disagreement Relative to Federated Anchor}

\subsection{Definition and Individual-Agent Bound}

Consistent with the revised manuscript, let $\V_G=\Psi(\Phi_G)$ denote the
decoded federated semantic anchor in the spatial task-demand space. Let
$\U_{\mathrm{act}}$ be a fixed active set over the analyzed calibration
sequence. Let fixed normalized aggregation weights over this sequence satisfy
\begin{equation}
\alpha_i\ge\alpha_{\min}>0,
\qquad
\sum_{i\in\U_{\mathrm{act}}}\alpha_i=1.
\end{equation}
For a fixed federated semantic anchor reference $\V_G$, define the
anchor-relative semantic-model disagreement as
\begin{equation}
\label{eq:supp_D_definition}
D_t
=
\sum_{i\in\U_{\mathrm{act}}}
\alpha_i
\|\V_i(t)-\V_G\|^2.
\end{equation}

Here, $\|\V_i(t)-\V_G\|$ quantifies disagreement directly in the decoded
spatial task-demand space, and is therefore used as the semantic-space
discrepancy rather than a neural-parameter-space distance. This choice is
consistent with the actionable representation consumed by the decentralized
policy for mission-aware spatial decisions.

\begin{proposition}[Individual Anchor-Relative Semantic-Model Discrepancy Bound]
For every $i\in\U_{\mathrm{act}}$,
\begin{equation}
\label{eq:supp_individual_D}
\|\V_i(t)-\V_G\|
\le
\sqrt{\frac{D_t}{\alpha_i}}
\le
\sqrt{\frac{D_t}{\alpha_{\min}}}.
\end{equation}
\end{proposition}

\begin{proof}
Because every term in~\eqref{eq:supp_D_definition} is nonnegative,
\begin{equation}
D_t
=
\sum_{j\in\U_{\mathrm{act}}}
\alpha_j
\|\V_j(t)-\V_G\|^2
\ge
\alpha_i
\|\V_i(t)-\V_G\|^2.
\end{equation}
Since $\alpha_i>0$,
\begin{equation}
\|\V_i(t)-\V_G\|^2
\le
\frac{D_t}{\alpha_i}.
\end{equation}
Taking square roots gives
\[
\|\V_i(t)-\V_G\|
\le
\sqrt{\frac{D_t}{\alpha_i}}.
\]
Finally, $\alpha_i\ge\alpha_{\min}$ implies
\[
\frac{1}{\alpha_i}
\le
\frac{1}{\alpha_{\min}},
\]
which gives~\eqref{eq:supp_individual_D}.
\end{proof}

\subsection{Calibration Contraction}

Let $\Delta_c\le T_{\mathrm{cal}}$ be the interval between successive
effective calibration events. For the contraction analysis below, the
federated semantic anchor reference $\V_G$ is conditioned to remain fixed
over the analyzed calibration sequence.

\begin{assumption}[Open-Loop Semantic-Model Disagreement Accumulation]
\label{assump:open_loop_disagreement}
Over each analyzed calibration interval, the pre-calibration disagreement
satisfies
\begin{equation}
\label{eq:supp_precal}
D_{(r+1)\Delta_c^-}
\le
D_{r\Delta_c}
+
C_D,
\qquad
C_D
=
c_1\Delta_c^2\epsilon_{\max}^2
+
c_2\varpi^2.
\end{equation}
The first term models the contribution of bounded spatial extrapolation
error accumulated during the open-loop interval, while the second term
accounts for bounded observation uncertainty. This is a sufficient
open-loop drift condition for the following contraction analysis; it is
not asserted as a general convergence result for asynchronous federated
learning.
\end{assumption}

At calibration epoch $t_r=(r+1)\Delta_c$, let $\V_i^-$ denote the
pre-calibration decoded local representation of participating active agent
$i$, and let $\V_G=\Psi(\Phi_G)$ denote the federated semantic anchor at the
analyzed calibration event. Consistent with~(23) of the revised manuscript,
the decoded representation is updated according to
\begin{equation}
\label{eq:supp_cal_update}
\V_i^+
=
(1-\omega_{i,t_r})\V_i^-
+
\omega_{i,t_r}\V_G.
\end{equation}
Assume
\begin{equation}
\omega_{i,t_r}\ge\omega_{\min}>0.
\end{equation}

\begin{proposition}[Bounded Anchor-Relative Semantic-Model Disagreement]\label{prop:anchor_disagreement}
Under the fixed-anchor condition,
\[
q=(1-\omega_{\min})^2\in[0,1),
\]
and
\begin{equation}
\label{eq:supp_D_recurrence}
D_{(r+1)\Delta_c}
\le
qD_{r\Delta_c}
+
qC_D.
\end{equation}
Consequently,
\begin{equation}
\label{eq:supp_D_limsup}
\limsup_{r\to\infty}D_{r\Delta_c}
\le
\frac{q}{1-q}
\left(
c_1\Delta_c^2\epsilon_{\max}^2
+
c_2\varpi^2
\right).
\end{equation}
\end{proposition}

\begin{proof}
Under the fixed-anchor condition, subtract $\V_G$ from~\eqref{eq:supp_cal_update}:
\begin{align}
\V_i^+-\V_G
&=
(1-\omega_{i,t_r})\V_i^-
+
\omega_{i,t_r}\V_G
-
\V_G
\nonumber\\
&=
(1-\omega_{i,t_r})(\V_i^--\V_G).
\end{align}
Therefore,
\begin{equation}
\|\V_i^+-\V_G\|^2
=
(1-\omega_{i,t_r})^2
\|\V_i^--\V_G\|^2.
\end{equation}
Since $0\le\omega_{i,t_r}\le1$,
\[
0\le1-\omega_{i,t_r}
\le
1-\omega_{\min},
\]
and hence
\begin{equation}
\|\V_i^+-\V_G\|^2
\le
(1-\omega_{\min})^2
\|\V_i^--\V_G\|^2
=
q\|\V_i^--\V_G\|^2.
\end{equation}
Multiplying by $\alpha_i$ and summing over the active set gives
\begin{align}
D_{(r+1)\Delta_c}
&=
\sum_{i\in\U_{\mathrm{act}}}
\alpha_i
\|\V_i^+-\V_G\|^2
\nonumber\\
&\le
q
\sum_{i\in\U_{\mathrm{act}}}
\alpha_i
\|\V_i^--\V_G\|^2
\nonumber\\
&=
qD_{(r+1)\Delta_c^-}.
\end{align}
Combining this with~\eqref{eq:supp_precal} yields
\[
D_{(r+1)\Delta_c}
\le
qD_{r\Delta_c}+qC_D,
\]
which is~\eqref{eq:supp_D_recurrence}.

We now unroll the recurrence. For one round,
\[
D_{\Delta_c}
\le
qD_0+qC_D.
\]
For two rounds,
\begin{align}
D_{2\Delta_c}
&\le
qD_{\Delta_c}+qC_D
\nonumber\\
&\le
q(qD_0+qC_D)+qC_D
\nonumber\\
&=
q^2D_0+qC_D(1+q).
\end{align}
After $R$ rounds,
\begin{equation}
D_{R\Delta_c}
\le
q^RD_0
+
qC_D
\sum_{j=0}^{R-1}q^j.
\end{equation}
Since $0\le q<1$,
\[
\sum_{j=0}^{R-1}q^j
=
\frac{1-q^R}{1-q},
\]
so
\begin{equation}
D_{R\Delta_c}
\le
q^RD_0
+
\frac{qC_D(1-q^R)}{1-q}.
\end{equation}
Taking $R\to\infty$ gives
\[
q^RD_0\to0,
\qquad
1-q^R\to1,
\]
and therefore
\[
\limsup_{r\to\infty}D_{r\Delta_c}
\le
\frac{qC_D}{1-q},
\]
which is~\eqref{eq:supp_D_limsup}.
\end{proof}

\subsection{Protocol-Induced Lower Bound on the Calibration Gain}

The dynamic calibration gain is
\begin{equation}
\label{eq:supp_omega}
\omega_t
=
1-\exp
\left(
-\lambda_{\mathrm{cal}}\Delta_{i,\mathrm{BS}}(t)
\right),
\qquad
\lambda_{\mathrm{cal}}>0.
\end{equation}
At an effective calibration event $t_r$, let
$\Delta_{i,\mathrm{BS}}(t_r^-)$ denote the accumulated downlink staleness
immediately prior to reception and the subsequent reset. Under the adopted
discrete-time protocol, $\Delta_{i,\mathrm{BS}}(t_r^-)\ge1$. Therefore, the
gain used for that calibration satisfies
\begin{align}
\omega_{i,t_r}
&=
1-\exp
\left(
-\lambda_{\mathrm{cal}}\Delta_{i,\mathrm{BS}}(t_r^-)
\right)
\nonumber\\
&\ge
1-\exp(-\lambda_{\mathrm{cal}})
\nonumber\\
&\triangleq
\omega_{\min}>0.
\end{align}
It follows that
\begin{align}
q
&=
(1-\omega_{\min})^2
\nonumber\\
&=
\left[
1-
\left(1-e^{-\lambda_{\mathrm{cal}}}\right)
\right]^2
\nonumber\\
&=
e^{-2\lambda_{\mathrm{cal}}}
<1.
\end{align}
Thus, at effective calibration events under the fixed-anchor condition,
the adopted gain rule with $\lambda_{\mathrm{cal}}>0$ yields the strict
contraction factor $q<1$.

\section{Cooperative Coverage Objective and State-Conditional Congestion Game}

\subsection{Coverage Objective and Fair-Share Utility}

Consistent with Definition~2 of the revised manuscript, the demand weights
$w_m(t)$, SGI values $\Gamma_m(t)$, and inferred peer positions are treated
as fixed exogenous components of $\mathbf z_t$ within the stage game.
Hence, $q_m(t)$ and the navigation guidance are fixed with respect to a
unilateral action deviation, except for the deviating agent's own
action-dependent navigation term. Unless otherwise stated, all
agent-indexed sums in this section are over $\U_{\mathrm{act}}$.

For a fixed state $\mathbf z_t$, define the occupancy of grid $m$ as
\begin{equation}
n_m(\mathbf a)
=
\sum_{j\in\U_{\mathrm{act}}}
\chi_{j,m}(\mathbf a),
\end{equation}
and define its prioritized service demand by
\begin{equation}
q_m(t)
=
w_m(t)\rho(\Gamma_m(t)).
\end{equation}
The cooperative coverage objective is
\begin{equation}
\label{eq:supp_W}
W(\mathbf a;\mathbf z_t)
=
\sum_{m\in\G}
q_m(t)
\mathbb{1}\{n_m(\mathbf a)>0\}.
\end{equation}

The task utility allocated to agent $i$ is
\begin{equation}
\label{eq:supp_task_U}
R_{i,t}^{\mathrm{task}}
=
\sum_{m\in\G}
\chi_{i,m}(\mathbf a)
\frac{q_m(t)}
{\max\{n_m(\mathbf a),1\}}.
\end{equation}
For every grid $m$ with $n_m>0$,
\begin{align}
\sum_{i\in\U_{\mathrm{act}}}
\chi_{i,m}(\mathbf a)
\frac{q_m(t)}{n_m(\mathbf a)}
&=
\frac{q_m(t)}{n_m(\mathbf a)}
\sum_{i\in\U_{\mathrm{act}}}
\chi_{i,m}(\mathbf a)
\nonumber\\
&=
\frac{q_m(t)}{n_m(\mathbf a)}
n_m(\mathbf a)
\nonumber\\
&=
q_m(t).
\end{align}
If $n_m=0$, both the summand in~\eqref{eq:supp_task_U} and the
corresponding indicator in~\eqref{eq:supp_W} are zero. Therefore,
\begin{equation}
\label{eq:supp_exact_decomp}
\sum_{i\in\U_{\mathrm{act}}}
R_{i,t}^{\mathrm{task}}
=
W(\mathbf a;\mathbf z_t).
\end{equation}

\subsection{Exact Potential Property}

Consistent with the revised manuscript, the stage-game utility retains the
action-dependent components of the composite reward. The inference-drift
regularizer is state-dependent and independent of agent $i$'s immediate
action at the current decision epoch; therefore, it does not affect
unilateral utility differences and can be omitted from the stage-game
utility without changing the exact-potential property.

For fixed $\mathbf z_t$, consider the stage utility
\begin{equation}
\label{eq:supp_U}
U_i(\mathbf a;\mathbf z_t)
=
\frac{1}{\zeta}
\left[
R_{i,t}^{\mathrm{task}}(\mathbf a)
+
R_{i,t}^{\mathrm{nav}}(\mathbf a_i;\mathbf z_t)
-
\lambda_{\mathrm{col}}
\sum_{j\ne i}
\mathbb I_{\mathrm{collision}}(i,j)
\right].
\end{equation}
The navigation guidance is treated as exogenous because $\mathbf z_t$ is
fixed; hence $R_{i,t}^{\mathrm{nav}}$ depends on $\mathbf a_i$ only.

Define the harmonic potential
\begin{equation}
\label{eq:supp_Phi}
\tilde\Phi_t(\mathbf a)
=
\frac{1}{\zeta}
\left[
\sum_{m\in\G}
q_m(t)H_{n_m(\mathbf a)}
+
\sum_{i\in\U_{\mathrm{act}}}R_{i,t}^{\mathrm{nav}}(\mathbf a_i;\mathbf z_t)
-
\frac12
\sum_{i\in\U_{\mathrm{act}}}
\sum_{\substack{j\in\U_{\mathrm{act}}\\j\ne i}}
\lambda_{\mathrm{col}}
\mathbb I_{\mathrm{collision}}(i,j)
\right],
\end{equation}
where
\[
H_n=\sum_{r=1}^{n}\frac1r,
\qquad
H_0=0.
\]

\begin{theorem}[Equilibrium Characterization of the State-Conditional Congestion Game]
\label{thm:supp_exact_potential}
For every fixed state $\mathbf z_t$, $\tilde\Phi_t$ in~\eqref{eq:supp_Phi}
is an exact potential function of the state-conditional stage game:
for every unilateral deviation $a_i\to a_i'$,
\begin{equation}
\label{eq:supp_exact_potential}
U_i(a_i',\mathbf a_{-i};\mathbf z_t)
-
U_i(a_i,\mathbf a_{-i};\mathbf z_t)
=
\tilde\Phi_t(a_i',\mathbf a_{-i})
-
\tilde\Phi_t(a_i,\mathbf a_{-i}).
\end{equation}
\end{theorem}

\begin{proof}
Fix $\mathbf a_{-i}$ and consider a unilateral deviation from $a_i$ to
$a_i'$. Define the set of abandoned cells
\begin{equation}
\mathcal M_i^-
=
\left\{
m:
\chi_{i,m}(a_i)=1,\;
\chi_{i,m}(a_i')=0
\right\},
\end{equation}
and the set of newly covered cells
\begin{equation}
\mathcal M_i^+
=
\left\{
m:
\chi_{i,m}(a_i)=0,\;
\chi_{i,m}(a_i')=1
\right\}.
\end{equation}

For a newly covered cell $m\in\mathcal M_i^+$, the occupancy changes from
$n_m$ to $n_m+1$. Its contribution to the task utility variation is
\[
\frac{q_m(t)}{n_m+1}.
\]
For an abandoned cell $m\in\mathcal M_i^-$, the occupancy changes from
$n_m$ to $n_m-1$, and its task-utility contribution decreases by
\[
\frac{q_m(t)}{n_m}.
\]
Thus
\begin{equation}
\label{eq:supp_task_delta}
\Delta R_{i,t}^{\mathrm{task}}
=
\sum_{m\in\mathcal M_i^+}
\frac{q_m(t)}{n_m+1}
-
\sum_{m\in\mathcal M_i^-}
\frac{q_m(t)}{n_m}.
\end{equation}

Now consider the harmonic part of the potential. For a newly covered
cell,
\begin{align}
q_m(t)
\left(
H_{n_m+1}-H_{n_m}
\right)
&=
q_m(t)\frac{1}{n_m+1}.
\end{align}
For an abandoned cell,
\begin{align}
q_m(t)
\left(
H_{n_m-1}-H_{n_m}
\right)
&=
-q_m(t)\frac{1}{n_m}.
\end{align}
Therefore,
\begin{align}
\Delta\tilde\Phi_t^{\mathrm{task}}
&=
\sum_{m\in\mathcal M_i^+}
\frac{q_m(t)}{n_m+1}
-
\sum_{m\in\mathcal M_i^-}
\frac{q_m(t)}{n_m}
\nonumber\\
&=
\Delta R_{i,t}^{\mathrm{task}}.
\label{eq:supp_harmonic_delta}
\end{align}

The navigation term is a function of $a_i$ only under fixed $\mathbf z_t$,
so its unilateral change is reproduced exactly by the corresponding
single-agent term in~\eqref{eq:supp_Phi}.

Finally, by the symmetric minimum-separation definition, the pairwise
collision indicator satisfies
\begin{equation}
\mathbb I_{\mathrm{collision}}(i,j)
=
\mathbb I_{\mathrm{collision}}(j,i).
\end{equation}
The double sum in~\eqref{eq:supp_Phi} counts each unordered pair twice,
hence the factor $1/2$ makes its variation equal to the collision term
seen by agent $i$ in~\eqref{eq:supp_U}.

Combining the task, navigation, and collision variations and using the
same positive normalization factor $\zeta$ in utility and potential gives
\[
\Delta U_i
=
\Delta\tilde\Phi_t,
\]
which proves~\eqref{eq:supp_exact_potential}.
\end{proof}

\subsection{Finite-Action Equilibrium Existence}

Consistent with the revised manuscript, consider a canonical discretization
of the operational action space $\A_i$, or equivalently a finite set of
navigational motion primitives, so that each $\A_i$ is finite. Since the
game has an exact potential, any strict unilateral improvement satisfies
\begin{equation}
U_i(a_i',\mathbf a_{-i};\mathbf z_t)
>
U_i(a_i,\mathbf a_{-i};\mathbf z_t)
\quad\Longrightarrow\quad
\tilde\Phi_t(a_i',\mathbf a_{-i})
>
\tilde\Phi_t(a_i,\mathbf a_{-i}).
\end{equation}
A strict better-response process therefore generates a strictly increasing
sequence of potential values. Because the joint action space
$\prod_i\A_i$ is finite, infinitely many strict improvements are impossible.
Hence the process terminates at an action profile with no strict unilateral
improvement, which is a pure-strategy Nash equilibrium (PNE).

Moreover, at a PNE $\mathbf a^*$,
\begin{equation}
\tilde\Phi_t(a_i',\mathbf a_{-i}^*)
\le
\tilde\Phi_t(\mathbf a^*)
\qquad
\forall i,\quad
\forall a_i'\in\A_i.
\end{equation}
Thus a PNE is a unilateral local maximum of the exact potential. This does
not imply global maximization of $W$ or Pareto optimality.

\section{Approximate Equilibrium under Intermittent Connectivity}

\subsection{Composite State Uncertainty}

Let $\hat{\mathbf z}_{i,t}$ denote the state perceived by agent $i$ under
intermittent connectivity. The state uncertainty is decomposed into
physical inference error, anchor-relative semantic-model disagreement, and
observation uncertainty.

\begin{assumption}[Composite State-Uncertainty Bound]
\label{assump:composite_uncertainty}
There exist constants $c_p,c_v,c_\varpi\ge0$ such that, for every active
agent $i$ over the analyzed outage horizon,
\begin{equation}
\label{eq:supp_composite}
\|\mathbf z_t-\hat{\mathbf z}_{i,t}\|
\le
c_p\epsilon_{T_{\mathrm{up}}}
+
c_v\sqrt{D_{T_{\mathrm{up}}}}
+
c_\varpi\varpi
\triangleq
\varepsilon_z(T_{\mathrm{up}}).
\end{equation}
The constants absorb unit conversions and local sensitivities that map the
three error components into the common state metric used by the stage utility.
\end{assumption}

By the individual anchor-relative semantic-model bound~\eqref{eq:supp_individual_D},
the semantic-model component for any active agent satisfies
\[
\|\V_i-\V_G\|
\le
\frac{1}{\sqrt{\alpha_{\min}}}\sqrt{D_t}.
\]
Consequently, the factor $1/\sqrt{\alpha_{\min}}$ can be absorbed into
the constant $c_v$ in~\eqref{eq:supp_composite}.

Lemma~\ref{lemma:drift} provides the finite-horizon physical contribution,
while Proposition~\ref{prop:anchor_disagreement} provides a bounded
anchor-relative semantic-model contribution under the fixed-anchor condition. Together with
Assumption~\ref{assump:composite_uncertainty}, the composite state mismatch
therefore remains finite over the analyzed finite horizon whenever
$\epsilon_{T_{\mathrm{up}}}$, $D_{T_{\mathrm{up}}}$, and $\varpi$ are finite.

\subsection{Coverage and Guidance Regularity}

Proposition~\ref{prop:semantic_error} confines a coverage-indicator mismatch
to the error-dependent boundary region characterized by
\eqref{eq:supp_semantic_condition}, which is contained in the conservative
$2\epsilon_k$ enclosing band described above. To remove the associated
binary discontinuities from the local utility-perturbation analysis,
impose the boundary-margin condition
\begin{equation}
\label{eq:supp_boundary_margin}
\epsilon_k<\mu_\partial,
\qquad
\mu_\partial>0,
\end{equation}
where all admissible state perturbations remain at least $\mu_\partial$
away from every coverage or safety boundary. Here,
$\epsilon_k<\mu_\partial$ is the single-position sufficient condition
relevant to the coverage classification in Proposition~\ref{prop:semantic_error}.
For the pairwise safety indicator, the admissible perturbation set is
additionally restricted so that the induced change in pairwise separation
remains smaller than its margin to $d_{\mathrm{safe}}$.

Under these boundary-margin conditions, a sufficiently small admissible
state perturbation cannot cross either type of boundary. Therefore, the
binary coverage indicators and minimum-separation violation indicators remain
locally invariant over the analyzed perturbation set. In particular,
\begin{equation}
\chi_{i,m}(\mathbf p)
=
\chi_{i,m}(\hat{\mathbf p})
\label{eq:supp_coverage_invariance}
\end{equation}
for every relevant cell, while the corresponding pairwise
$\mathbb I_{\mathrm{collision}}(i,j)$ has the same binary value under the
true and perceived states for every active-agent pair.

Next, let $m_i^*$ denote the highest-deficit hotspot selected by agent $i$.
The positive task-direction margin assumption makes this maximizer unique
over the admissible perturbation set. We additionally require the agent
position to remain a positive distance from this hotspot:
\begin{equation}
\label{eq:supp_hotspot_distance}
\|\mathbf p_i-\mathbf g_{m_i^*}\|>0.
\end{equation}
The normalized guidance direction is
\begin{equation}
\mathbf d_i
=
\frac{\mathbf g_{m_i^*}-\mathbf p_i}
{\|\mathbf g_{m_i^*}-\mathbf p_i\|}.
\end{equation}
On any compact region satisfying
$\|\mathbf p_i-\mathbf g_{m_i^*}\|\ge d_{\min}>0$, this mapping is locally
Lipschitz. Indeed, for
\[
\mathbf x,\mathbf y\ne\mathbf0,
\]
the normalized-vector map $f(\mathbf x)=\mathbf x/\|\mathbf x\|$ satisfies
\begin{equation}
\left\|
\frac{\mathbf x}{\|\mathbf x\|}
-
\frac{\mathbf y}{\|\mathbf y\|}
\right\|
\le
\frac{2}{d_{\min}}\|\mathbf x-\mathbf y\|
\end{equation}
whenever $\|\mathbf x\|,\|\mathbf y\|\ge d_{\min}$.
Therefore the navigation direction contributes a finite Lipschitz term
to the utility perturbation.

\subsection{Uniform Stage-Utility Perturbation}

Define the perceived stage utility as
\begin{equation}
\hat U_i(\mathbf a)
\triangleq
U_i(\mathbf a;\hat{\mathbf z}_{i,t}).
\end{equation}
Under the boundary margin, the coverage-indicator invariance in
\eqref{eq:supp_coverage_invariance}, the corresponding minimum-separation
status invariance, unique hotspot selection, positive hotspot distance, and
local Lipschitz continuity of the decoder $\Psi$, all remaining state-dependent components
of the stage utility are locally Lipschitz over the resulting compact
operational region.

Let $\bar L_U<\infty$ denote a common Lipschitz modulus with respect to the
composite state uncertainty metric. Then
\begin{align}
\left|
U_i(\mathbf a;\mathbf z_t)
-
\hat U_i(\mathbf a)
\right|
&=
\left|
U_i(\mathbf a;\mathbf z_t)
-
U_i(\mathbf a;\hat{\mathbf z}_{i,t})
\right|
\nonumber\\
&\le
\bar L_U
\|\mathbf z_t-\hat{\mathbf z}_{i,t}\|
\nonumber\\
&\le
\bar L_U
\varepsilon_z(T_{\mathrm{up}}).
\end{align}
Taking the supremum over the admissible action set gives
\begin{equation}
\label{eq:supp_lipschitz_bridge}
\sup_{\mathbf a\in\A}
\left|
U_i(\mathbf a;\mathbf z_t)
-
\hat U_i(\mathbf a)
\right|
\le
\delta_U(T_{\mathrm{up}})
\le
\bar L_U\varepsilon_z(T_{\mathrm{up}}).
\end{equation}

The use of a uniform bound is deliberate: the subsequent
$\epsilon$-Nash transfer compares the equilibrium action with every
admissible unilateral alternative.

\subsection{\texorpdfstring{$\epsilon$-Nash Robustness}{epsilon-Nash Robustness}}

The following perturbation-transfer result is conditional on the existence
of a pure-strategy equilibrium of the perceived stage game. It does not
assert that heterogeneous perceived states preserve the exact-potential
structure of the complete-information state-conditional game.

\begin{corollary}[$\epsilon$-Nash Equilibrium under Bounded Estimation Error]
\label{cor:supp_epsilon_ne}
Let $\hat{\mathbf a}^*$ be a pure-strategy Nash equilibrium of the perceived
stage game $\hat{\mathcal G}_t$, and suppose~\eqref{eq:supp_lipschitz_bridge}
holds. Then $\hat{\mathbf a}^*$ is an
$\epsilon_{\mathrm{NE}}$-approximate Nash equilibrium of the true
stage game, with
\begin{equation}
\label{eq:supp_epsilon_ne}
\epsilon_{\mathrm{NE}}
\le
2\delta_U(T_{\mathrm{up}})
\le
2\bar L_U\varepsilon_z(T_{\mathrm{up}}).
\end{equation}
\end{corollary}

\begin{proof}
Because $\hat{\mathbf a}^*$ is a Nash equilibrium of the perceived game,
for every active agent $i$ and every unilateral alternative
$a_i'\in\A_i$,
\begin{equation}
\hat U_i(\hat{\mathbf a}^*)
\ge
\hat U_i(a_i',\hat{\mathbf a}_{-i}^*).
\label{eq:supp_perceived_NE}
\end{equation}
From the uniform perturbation bound,
\begin{equation}
U_i(\hat{\mathbf a}^*;\mathbf z_t)
\ge
\hat U_i(\hat{\mathbf a}^*)
-
\delta_U.
\label{eq:supp_true_lower}
\end{equation}
Combining~\eqref{eq:supp_true_lower} with~\eqref{eq:supp_perceived_NE},
\begin{align}
U_i(\hat{\mathbf a}^*;\mathbf z_t)
&\ge
\hat U_i(a_i',\hat{\mathbf a}_{-i}^*)
-
\delta_U.
\end{align}
Applying the perturbation bound once more at the unilateral alternative,
\begin{equation}
\hat U_i(a_i',\hat{\mathbf a}_{-i}^*)
\ge
U_i(a_i',\hat{\mathbf a}_{-i}^*;\mathbf z_t)
-
\delta_U.
\end{equation}
Therefore,
\begin{equation}
U_i(\hat{\mathbf a}^*;\mathbf z_t)
\ge
U_i(a_i',\hat{\mathbf a}_{-i}^*;\mathbf z_t)
-
2\delta_U.
\end{equation}
Since $a_i'$ is arbitrary,
\begin{equation}
\max_{a_i'\in\A_i}
\left[
U_i(a_i',\hat{\mathbf a}_{-i}^*;\mathbf z_t)
-
U_i(\hat{\mathbf a}^*;\mathbf z_t)
\right]
\le
2\delta_U.
\end{equation}
Taking the maximum over all active agents gives
\begin{equation}
\operatorname{Regret}(\hat{\mathbf a}^*)
\le
2\delta_U(T_{\mathrm{up}}).
\end{equation}
Using~\eqref{eq:supp_lipschitz_bridge},
\[
\operatorname{Regret}(\hat{\mathbf a}^*)
\le
2\bar L_U\varepsilon_z(T_{\mathrm{up}}),
\]
which proves~\eqref{eq:supp_epsilon_ne}.
\end{proof}

\section{Combined Analytical Chain}

The preceding results establish the same analytical progression as Section~V of the revised manuscript.

First, the \emph{Kinematic Error and Semantic Robustness} analysis uses Lemma~\ref{lemma:drift} to obtain the finite-horizon physical inference bound
\begin{equation}
\epsilon_k
\le
\delta t
\sum_{\ell=0}^{k-1}
\left[
\|\mathbf{v}_t-\hat{\mathbf{v}}_t\|
+
a_{\max}\ell\delta t
+
\|\hat{\mathbf{v}}_t\|
\left(1-(1-\alpha)^\ell\right)
\right],
\end{equation}
and Proposition~\ref{prop:semantic_error} maps this physical error into the semantic coverage-misclassification condition
\begin{equation}
\chi_{i,m}(\mathbf p)\neq\chi_{i,m}(\hat{\mathbf p})
\quad\Longrightarrow\quad
\left|
\|\hat{\mathbf p}-\mathbf g_m\|-R_{\mathrm{cov}}
\right|
\le\epsilon_k.
\end{equation}

Second, the \emph{Bounded Semantic-Model Disagreement Relative to Federated Anchor} analysis characterizes federated semantic-model alignment through the anchor-relative disagreement $D_t$ in the decoded spatial task-demand space. Using $\alpha_i\ge\alpha_{\min}>0$ gives the individual-agent bound
\begin{equation}
\|\V_i-\V_G\|
\le
\frac{1}{\sqrt{\alpha_{\min}}}\sqrt{D_t}.
\end{equation}

Third, the \emph{Cooperative Coverage Objective and State-Conditional Congestion Game} analysis establishes the complete-information equilibrium reference. Theorem~\ref{thm:supp_exact_potential} shows that, for every fixed $\mathbf z_t$, the adopted state-conditional congestion game admits a Rosenthal-type exact potential; under finite action sets, it therefore admits at least one PNE. This result characterizes the complete-information stage game and does not assert that heterogeneous perceived states preserve the same exact-potential structure.

Finally, the \emph{Approximate Equilibrium under Intermittent Connectivity} analysis combines these ingredients. The composite uncertainty bound~\eqref{eq:supp_composite} converts physical inference error, anchor-relative semantic-model disagreement, and observation uncertainty into a bounded state perturbation. The local Lipschitz bridge then yields the uniform stage-utility perturbation bound
\begin{equation}
\sup_{\mathbf a\in\A}
|U_i(\mathbf a;\mathbf z_t)-\hat U_i(\mathbf a)|
\le
\bar L_U\varepsilon_z(T_{\mathrm{up}}),
\end{equation}
which, through Corollary~\ref{cor:supp_epsilon_ne}, gives
\begin{equation}
\epsilon_{\mathrm{NE}}
\le
2\bar L_U\varepsilon_z(T_{\mathrm{up}}).
\end{equation}

Thus, the robustness statement remains a finite-horizon perturbation result: as physical inference error, anchor-relative semantic-model disagreement, and observation uncertainty decrease, the maximum unilateral regret of a perceived-game equilibrium decreases proportionally under the stated regularity conditions.